\documentclass[aps,pre,twocolumn,amssymb,amsfonts,floatfix,superscriptaddress,longbibliography,notitlepage,footinbib]{revtex4-1}
\usepackage[utf8]{inputenc}
\pdfoutput=1
\usepackage{color}
\usepackage{float}
\usepackage{xcolor}
\usepackage{mathtools,amsmath,amsthm,amsfonts,amssymb}
\usepackage{commath}
\usepackage{physics}
\usepackage{bm}
\usepackage[pdftex]{graphicx}
\usepackage[colorlinks=true,linkcolor=blue,
            citecolor=blue,
            urlcolor=blue]{hyperref}
\usepackage{enumitem}
\usepackage{soul}

\begin{document}

\title{Correspondence principle, dissipation, and Ginibre ensemble}

\author{David Villase\~nor}
\author{Hua Yan}
\author{Matic Orel}
\author{Marko Robnik} 
\affiliation{CAMTP - Center for Applied Mathematics and Theoretical Physics, University of Maribor, Mladinska 3, SI-2000 Maribor, Slovenia, European Union}


\begin{abstract}
The correspondence between quantum and classical behavior has been essential since the advent of quantum mechanics. This principle serves as a cornerstone for understanding quantum chaos, which has garnered increased attention due to its strong impact in various theoretical and experimental fields. When dissipation is considered, quantum chaos takes concepts from isolated quantum chaos to link classical chaotic motion with spectral correlations of Ginibre ensembles. This correspondence was first identified in periodically kicked systems with damping, but it has been shown to break down in dissipative atom-photon systems~[\href{https://journals.aps.org/prl/abstract/10.1103/PhysRevLett.133.240404}{Phys. Rev. Lett. 133, 240404 (2024)}]. In this contribution, we revisit the original kicked model and perform a systematic exploration across a broad parameter space, reaching a genuine semiclassical limit. Our results demonstrate that the correspondence principle, as defined through this spectral connection, fails even in this prototypical system. These findings provide conclusive evidence that Ginibre spectral correlations are neither a robust nor a universal diagnostic of dissipative quantum chaos.
\end{abstract}

\maketitle

Quantum theory has raised a series of intriguing questions since its early development. One key idea is how quantum mechanics aligns with the predictions of classical mechanics, which led to the formulation of the correspondence principle~\cite{Bohr1920,Hassoun1989}. The systematic observation of chaotic motion in nonlinear systems~\cite{Lorenz1963}, along with its implications for engineering and complex systems~\cite{OttBook,Guckenheimer1983book,AnishchenkoBook,Shivamoggi2014book}, has sparked a growing interest in understanding how this phenomenon translates into quantum mechanics, giving rise to the fruitful field of quantum chaos~\cite{HaakeBook,CasatiBook,StockmannBook,WimbergerBook}. The impact of quantum chaos on both theoretical and experimental topics of current interest, such as thermalization~\cite{Borgonovi2016,Dalessio2016}, quantum metrology~\cite{Fiderer2018}, quantum information~\cite{Swingle2016,Landsman2019,Lantagne2020,Anand2021}, and high-energy physics~\cite{Jensen2016,Maldacena2016JHEP,Magan2018,Jahnke2019}, among others~\cite{Suntajs2020}, has motivated the study of dissipative quantum systems~\cite{BreuerBook,CarmichaelBook1993,CarmichaelBook2002}, where the chaotic motion in the classical limit is characterized by the emergence of chaotic attractors~\cite{Ott1981,Eckmann1985,Ku1990,Smyrlis1991,Papageorgiou1991}.

Traditionally, the correspondence principle links classical chaotic motion in isolated systems with universal spectral correlations, as suggested by \emph{the quantum chaos conjecture}~\cite{Casati1980,Bohigas1984}. In contrast, classical regular motion corresponds to uncorrelated energy spectra~\cite{Berry1977}. A key advancement in formulating a correspondence principle in dissipative systems involved extending the correspondence principle established in isolated systems to the periodically kicked top with damping~\cite{Grobe1988}. These studies demonstrated that the presence of a chaotic attractor in the classical limit corresponds to cubic level repulsion in the Liouvillian spectrum of a Floquet operator with complex phases, similar to what is seen in Ginibre ensembles of non-Hermitian Gaussian random matrices~\cite{Ginibre1965}. In contrast, simple attractors correspond to linear level repulsion in the Liouvillian spectrum. These two relationships were recognized as \emph{the dissipative quantum chaos conjecture}~\cite{Grobe1988,Grobe1989,Akemann2019}.

The description established by the dissipative conjecture has inspired the application of the spectral analysis to define quantum chaos in dissipative many-body quantum systems~\cite{Akemann2019,Jaiswal2019,Sa2020,Rubio2022,Garcia2023,Akemann2025}, which lack a classical limit, and categorize these systems based on their symmetries~\cite{Hamazaki2020,Garcia2022,Kawabata2023,Sa2023}. Subsequent studies have been conducted to pinpoint specific behaviors of the Ginibre spectral correlations in quantum evolution~\cite{Li2021,Shivam2023,Li2024ARXIV}, while others have attempted to detect these correlations experimentally~\cite{Wold2025Arxiv}. However, dissipative atom-field systems with a well-defined classical limit have detected that the description of the dissipative conjecture fails~\cite{Villasenor2024}, raising questions about the reliability of its predictions. Alternative frameworks for understanding dissipative chaos have also been proposed, using properties of steady states~\cite{Sa2020JPA,Sa2020PRB,Costa2023,Richter2025,Ferrari2025,Peyruchat2025,Mondal2025ARXIV,Rufo2025Arxiv,Li2025Arxiv} or focusing on dynamical signatures~\cite{Zhang2019,Garcia2024,Mondal2025ARXIV}.

The previous discussion provides a perspective on the perplexing nature of what dissipative quantum chaos entails. In this work, we investigate the periodically kicked top with dissipation~\cite{Grobe1988} that inspired the formulation of the dissipative quantum chaos conjecture. A variety of research has been conducted in this system, including studies of quantum dynamics~\cite{Grobe1987,Peplowski1991}, classical motion in presence of dissipation and pumping~\cite{Peplowski1988}, semiclassical analysis~\cite{Braun1998a,Braun1998b,Braun1999,Braun1999PhysD}, and alternative methods for modeling dissipation, such as stochastic methods~\cite{Iwaniszewski1995,Gisin1992,Spiller1994,Brun1996} and non-Hermitian Hamiltonian approaches~\cite{Mudute2020,Chalker1998,MoiseyevBook2011,Kawabata2019,Ashida2020,Akemann2022,Akemann2025Arxiv}. The characteristics of non-Hermitian Hamiltonians have led to useful correspondence principles between quantum and classical dynamics~\cite{Graefe2008,Graefe2010,Graefe2010JPA,Holmes2023}. The latest studies of the kicked top have shown a connection between dissipative classical chaos and measures of quantum complexity~\cite{Passarelli2025}. Moreover, the kicked top in absence of dissipation is a paradigm of classical and quantum chaos~\cite{Frahm1985,Frahm1986,Haake1987,Kus1987,Kus1988,Alicki1996,Constantoudis1997,Sreeram2025} and has been the subject of fruitful research, including mixed-type phase space dynamics~\cite{Wang2023b,Yan2024} under the Berry-Robnik picture~\cite{Berry1984,Robnik1998,Robnik2023},
scarring~\cite{Zhang1990,Kus1991,Dariano1992}, entanglement~\cite{Wang2004,Ghose2008,Chaudhury2009,Lombardi2011,Ruebeck2017,Piga2019,Kumari2019,Lerose2020,Zou2022,Khalid2025,Manju2025ARXIV}, localization~\cite{Wang2023a}, fractality~\cite{Nakamura1986,Wang2021,Gonzalez2025}, as well as experimental realizations~\cite{Smith2004,Smith2006,Chaudhury2007,Ghose2008,Chaudhury2009} and metrological aspects~\cite{Zou2025}.

Our analysis of the kicked top with dissipation examines a broad parameter space and reveals that the expected correspondence between quantum and classical behavior is not entirely valid. This provides conclusive evidence of the limitations of Ginibre spectral correlations as an overall test of dissipative quantum chaos.

{\it Quantum kicked top.}
A periodically kicked top with dissipation~\cite{Grobe1987,Grobe1988} is represented with the Markovian Lindblad master equation~\cite{BreuerBook,CarmichaelBook1993,CarmichaelBook2002}, setting $\hbar=1$,
\begin{equation}
    \label{eq:Lindblad}
    \frac{d\hat{\rho}}{dt} = \hat{\mathcal{L}}(t)\hat{\rho} = -i[\hat{H}(t),\hat{\rho}] + \hat{\Lambda}\hat{\rho} = \left(\hat{\Lambda} - i \hat{L}(t)\right)\hat{\rho} ,
\end{equation}
where $\hat{\mathcal{L}}(t)$ is the Liouvillian operator, $\hat{L}(t)\hat{\rho} = [\hat{H}(t),\hat{\rho}] = \hat{L}_{0}\hat{\rho} + \hat{L}_{1}\hat{\rho}\sum_{n}\delta(t-n)$, and the dissipation process is chosen as occurs in the theory of superradiance~\cite{Bonifacio1971}
\begin{equation}
    \hat{\Lambda}\hat{\rho} = \frac{\Gamma}{2j}\left(2\hat{J}_{-}\hat{\rho}\hat{J}_{+} - \{\hat{J}_{+}\hat{J}_{-},\hat{\rho}\}\right) ,
\end{equation}
where $\Gamma$ is the dissipation strength. The terms $\hat{L}_{0,1}\hat{\rho} = [\hat{H}_{0,1},\hat{\rho}]$ come from the unitary dynamics represented by the time-dependent Hamiltonian
\begin{equation}
    \label{eq:KTHamiltonian}
    \hat{H}(t) = \hat{H}_{0} + \hat{H}_{1}\sum_{n}\delta(t-n),
\end{equation}
where $\hat{H}_{0} = p\hat{J}_{z} + (k_{0}/(2j))\hat{J}_{z}^{2}$ is a precessional motion around the $z$ axis controlled by the parameters $p$ and $k_{0}$. The term $\hat{H}_{1} = (k_{1}/(2j))\hat{J}_{y}^{2}$ is a rotational motion around the $y$ axis that appears periodically with unit period and is modulated by the kick strength $k_{1}$. The Dirac delta function in Eq.~\eqref{eq:KTHamiltonian} represents the periodic kick. The operators $\hat{J}_{x,y,z}$ and $\hat{J}_{\pm} = \hat{J}_{x} \pm i \hat{J}_{y}$ are usual angular momentum operators.

The operator $\hat{\mathbf{J}}^{2}$ with eigenvalues $j(j+1)$ is a conserved quantity due to the commutation relations with the Hamiltonian in Eq.~\eqref{eq:KTHamiltonian}, the Floquet operator $\hat{F} = e^{-i\hat{H}_{1}}e^{-i\hat{H}_{0}}$, and the Liouvillian in Eq.~\eqref{eq:Lindblad}, $[\hat{H}(t),\hat{\mathbf{J}}^{2}]=[\hat{F},\hat{\mathbf{J}}^{2}]=[\hat{\mathcal{L}}(t),\hat{\mathbf{J}}^{2}]=0$. The angular momentum basis $|j,m_{z}\rangle$ provides a Hilbert space with finite dimension $2j+1$ and a Liouville space with dimension $(2j+1)^{2}$. The parity symmetry of the Hamiltonian system appears as a weak symmetry~\cite{Albert2014,Lieu2020} in the dissipative system, $[\hat{D},\hat{\mathcal{P}}]=[\hat{F},\hat{\Pi}]=0$, where $\hat{\Pi} = e^{i\pi(\hat{J}_{z} + j\hat{\mathbb{I}})}$ is the parity operator, $\hat{\mathcal{P}}\hat{\rho} = \hat{\Pi}\hat{\rho}\hat{\Pi}^{\dagger}$, and $\hat{D}$ is the dissipative Floquet operator that provides a kick-to-kick stroboscopic description of the density operator
\begin{equation}
    \label{eq:FloquetOperator}
    \hat{\rho}_{n+1} = \hat{D}\hat{\rho}_{n} = e^{- i \hat{L}_{1}}e^{\left(\hat{\Lambda} - i \hat{L}_{0}\right)} \hat{\rho}_{n} .
\end{equation}

{\it Classical limit of the kicked top.}
The classical limit of the periodically kicked top with dissipation is obtained by a mean-field approximation for the expectation values of the angular momentum operators $\langle \hat{J}_{x,y,z}\rangle = \text{Tr}(\hat{\rho}\hat{J}_{x,y,z})$, 
\begin{align}
    \label{eq:JxClassical}
    \dot{J}_{x} = & -\left[p + k_{0} J_{z} - k_{1} J_{z}\sum_{n}\delta(t-n)\right] J_{y} + \Gamma J_{x}J_{z} ,  \\
    \label{eq:JyClassical}
    \dot{J}_{y} = & \left(p + k_{0} J_{z}\right) J_{x} + \Gamma J_{y}J_{z} , \\
    \label{eq:JzClassical}
    \dot{J}_{z} = & -k_{1} J_{x}J_{y}\sum_{n}\delta(t-n) - \Gamma \left(J_{x}^{2} + J_{y}^{2}\right) ,
\end{align}
where we used the evolution law $\partial_{t}\hat{J}_{x,y,z} = \hat{\mathcal{L}}^{\dagger}(t)\hat{J}_{x,y,z}$ and decoupled expectation values $\langle\hat{J}_{x}\hat{J}_{y}\rangle \approx \langle\hat{J}_{x}\rangle\langle\hat{J}_{y}\rangle$. The classical variables $J_{x,y,z} = \lim_{j\to\infty}\langle \hat{J}_{x,y,z}\rangle / j$  represent an angular momentum vector in the Bloch sphere, $\mathbf{J}=(J_{x},J_{y},J_{z}) = (\cos\phi \sin\theta,\sin\phi \sin\theta,\cos\theta)$, with conserved unit magnitude $|\mathbf{J}|=1$.

The Hamiltonian dynamics can be recovered by switching off the dissipation strength $\Gamma$ in Eqs.~\eqref{eq:JxClassical}-\eqref{eq:JzClassical}, or by a mean-field approximation for the expectation value of the quantum Hamiltonian in Eq.~\eqref{eq:KTHamiltonian} with Bloch coherent states of minimum uncertainty, $|\alpha\rangle = e^{(\alpha\hat{J}_{+}-\alpha^{\ast}\hat{J}_{-})}|j,-j\rangle$, obtaining the classical Hamiltonian
\begin{equation}
    H(t) = \lim_{j\to\infty} \frac{\langle \alpha|\hat{H}(t)|\alpha\rangle}{j} = H_{0} + H_{1}\sum_{n}\delta(t-n) ,
\end{equation}
where $H_{0} = p J_{z} + (k_{0}/2)J_{z}^{2}$ and $H_{1} = (k_{1}/2)J_{y}^{2}$. The classical variables $J_{x,y,z} = \lim_{j\to\infty}\langle \alpha|\hat{J}_{x,y,z}|\alpha\rangle / j$ provide the Hamilton equations of motion $\dot{J}_{x,y,z} = \{ J_{x,y,z} , H(t) \}$.

The nonautonomous dynamical system in Eqs.~\eqref{eq:JxClassical}-\eqref{eq:JzClassical} represents two processes of the classical motion. In the first process, dissipation takes place continuously before the kick and we can set $k_{1}=0$ in Eqs.~\eqref{eq:JxClassical}-\eqref{eq:JzClassical}. The evolution of this reduced dynamical system provides the solution $J_{x,y,z}$. The second process is discrete and is defined by the action of the kick ($k_{1} \neq 0$) as a rotation around the $y$ axis of the previous solution, $\widetilde{J}_{x,y,z} = R(k_{1} J_{y})J_{x,y,z}$. The last two processes allow us to define a kick-to-kick stroboscopic description for the dissipative classical motion in the same spirit as occurs with the Hamiltonian motion
\begin{align}
    \label{eq:JxClassicalK}
    \widetilde{J}_{x}^{n+1} = & \cos(k_{1} J_{y}^{n})J_{x}^{n} + \sin(k_{1} J_{y}^{n})J_{z}^{n} , \\
    \label{eq:JyClassicalK}
    \widetilde{J}_{y}^{n+1} = & J_{y}^{n} , \\
    \label{eq:JzClassicalK}
    \widetilde{J}_{z}^{n+1} = & -\sin(k_{1} J_{y}^{n})J_{x}^{n} + \cos(k_{1} J_{y}^{n})J_{z}^{n} .
\end{align}
In the Supplemental Material~\cite{footSM}, we present an alternative classical description that considers dissipation as a decoupled process from the unitary evolution~\cite{Grobe1987,Braun1999,Braun1999PhysD}.

{\it Classical and quantum dissipative chaos.}
Chaotic motion in dissipative classical systems is described by the appearance of chaotic attractors through different routes, such as period-doubling cascades, intermittency, or quasiperiodicity~\cite{OttBook,Guckenheimer1983book,AnishchenkoBook,Shivamoggi2014book,Ott1981,Eckmann1985,Ku1990,Smyrlis1991,Papageorgiou1991}. The motion inside the chaotic attractors generates an exponential divergence of the initial conditions, which can be identified with Lyapunov exponents $ h = \lim_{t\to\infty}\lim_{\delta_{0}\to0} t^{-1}\ln\left(\delta(t)/\delta_{0}\right)$~\cite{Benettin1980a,Benettin1980b,Wolf1985}, where $\delta(t)=|\mathbf{x}_{1}(t)-\mathbf{x}_{2}(t)|$ is the separation in phase space of two trajectories at time $t$ and $\delta_{0}=\delta(0)$.

The dissipative quantum chaos conjecture relates the presence of chaotic attractors in the classical dynamics with a level spacing distribution described by the Ginibre unitary ensemble (GinUE) distribution~\cite{Ginibre1965,Grobe1988,Akemann2019}
\begin{equation}
    \label{eq:GinUEDistribution}
    P_{\text{GinUE}}(s) = \prod_{k=1}^{\infty}\frac{\Gamma(1+k,s^{2})}{k!}\sum_{k'=1}^{\infty}\frac{2s^{2k'+1}e^{-s^{2}}}{\Gamma(1+k',s^{2})} ,
\end{equation}
where $\Gamma(k,z) = \int_{z}^{\infty}dt\,t^{k-1}e^{-t}$ is the incomplete Gamma function. The spacing $s_{k}=|\varphi_{k}^{\text{NN}}-\varphi_{k}|$ is defined as the minimum Euclidean distance between a reference eigenvalue $\varphi_{k}$ and its nearest neighbor $\varphi_{k}^{\text{NN}}$ in the complex plane. On the contrary, when simple attractors appear in the classical dynamics, the spacing distribution agrees with a 2D Poisson function~\cite{Grobe1988,Akemann2019}
\begin{equation}
    \label{eq:2DPDistribution}
    P_{\text{2DP}}(s) = \frac{\pi}{2}s\,e^{-\pi s^{2}/4}.
\end{equation}
The GinUE distribution is normalized with a first moment given by $\bar{s} = \int_{0}^{\infty} ds\,s\,P_{\text{GinUE}}(s)\approx1.1429$. For numerical comparisons, we scale the GinUE distribution, $\widetilde{P}_{\text{GinUE}}(s) = \bar{s}P_{\text{GinUE}}(\bar{s}s)$, to ensure that its first moment is also normalized, $\int_{0}^{\infty} ds\,s\,\widetilde{P}_{\text{GinUE}}(s)=1$. The limit $s\to0$ provides the power of level repulsion, $P_{\beta}(s) \propto s^{\beta}$, which is determined by the exponent $\beta$. In this way, linear (cubic) level repulsion, $\beta=1$ ($\beta=3$), identifies a regular (chaotic) dissipative quantum system~\cite{HaakeBook,Grobe1988,Grobe1989}.

{\it Quantum analysis of the kicked top.}
We begin with the spectral analysis of the kicked top with dissipation by considering the complex eigenphases $\varphi_{k}$ of the dissipative Floquet operator defined in Eq.~\eqref{eq:FloquetOperator}. The eigenphases are given by the eigenvalue equation $\hat{D}|\lambda_{k}\rangle\rangle=\lambda_{k}|\lambda_{k}\rangle\rangle=e^{\varphi_{k}}|\lambda_{k}\rangle\rangle$. As occurs in isolated systems, the complex eigenphases need to be unfolded to remove system specific behaviors~\cite{Markum1999,Akemann2019,Hamazaki2020}. To avoid this procedure, we use the complex spectral ratio introduced in Ref.~\cite{Sa2020}:
\begin{equation}
    \label{eq:ComplexRatio}
    Z_{k}=r_{k}e^{i\theta_{k}}=\frac{\varphi_{k}^{\text{NN}}-\varphi_{k}}{\varphi_{k}^{\text{NNN}}-\varphi_{k}} ,
\end{equation}
where NN stands for the nearest neighbor and NNN for the next-to-nearest neighbor of the reference eigenphase $\varphi_{k}$. The average of the quantities $r_{k}=|Z_{k}|$ and $\theta_{k}=\text{Arg}(Z_{k})$ provide limit values for numerical comparisons. The averages $\langle r\rangle_{\text{2DP}} = 2/3$  and $-\langle \cos\theta\rangle_{\text{2DP}} = 0$ are obtained using the 2D Poisson distribution. For the GinUE distribution, we get $\langle r\rangle_{\text{GinUE}} \approx 0.74$ and $-\langle \cos\theta\rangle_{\text{GinUE}} \approx 0.24$.

\begin{figure}[ht]
\centering
\includegraphics[width=\columnwidth]{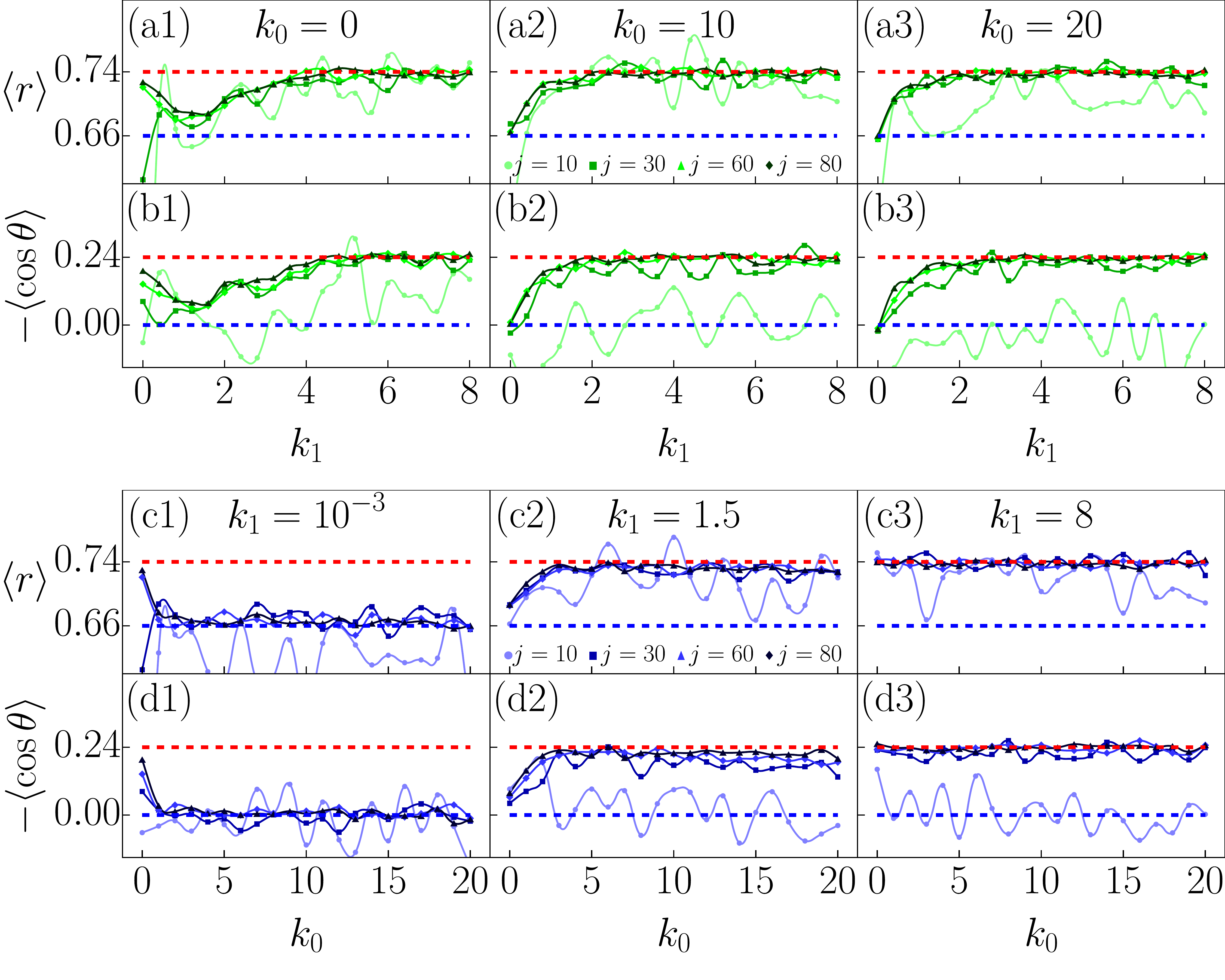}
\caption{System size analysis in the kicked top with dissipation. [(a1)-(a3)] Average $\langle r \rangle$ and [(b1)-(b3)] average $-\langle \cos\theta \rangle$ from the complex spectral ratio in Eq.~\eqref{eq:ComplexRatio} as a function of the kick strength $k_{1}$. Each column identifies a different case: $k_{0}=0$, $k_{0}=10$, and $k_{0}=20$. [(c1)-(c3)] Average $\langle r \rangle$ and [(d1)-(d3)] average $-\langle \cos\theta \rangle$ as a function of the parameter $k_{0}$. Each column identifies a different case: $k_{1}=10^{-3}$, $k_{1}=1.5$, and $k_{1}=8$. In all panels, we show the last averages for different system sizes $j=10,30,60,80$. The horizontal blue (red) dashed line represents the limit of a quantum system described by 2D Poisson (GinUE) statistics. System parameters: $p=2$ and $\Gamma=0.1$.}
\label{fig:SystemSize}
\end{figure}

Next, we present the analysis of the complex spectral ratio for the kicked top with dissipation and use the parameters considered in Ref.~\cite{Grobe1988}. We restrict our study to the dissipation region $0<\Gamma\leq0.4$. For dissipation strengths $\Gamma>0.4$, the numerical eigenvalues $\lambda_{k}$ collapse to the center of the unit circle and the corresponding eigenphases $\varphi_{k}$ are basically random numbers below the machine precision (see the Supplemental Material~\cite{footSM} for complementary information).

\begin{figure*}[ht]
\centering  
\includegraphics[width=\textwidth]{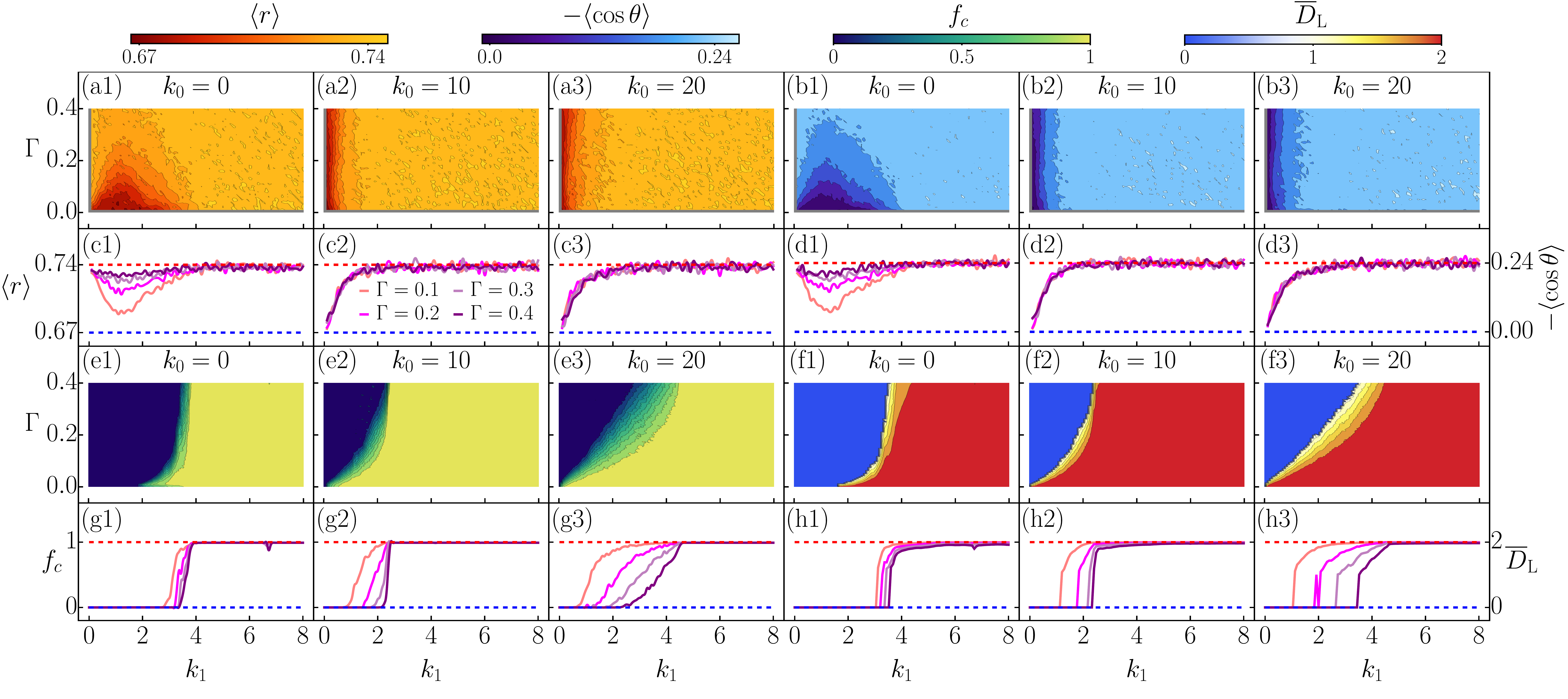}
\caption{Quantum and classical analysis of the kicked top with dissipation. [(a1)-(a3)] Average $\langle r \rangle$ and [(b1)-(b3)] average $-\langle \cos\theta \rangle$ from the complex spectral ratio in Eq.~\eqref{eq:ComplexRatio} as a function of the kick and dissipation strengths ($k_{1}$,$\Gamma$). Panels (c1)-(c3) and (d1)-(d3) show the last averages for specific dissipation strengths, $\Gamma=0.1,0.2,0.3,0.4$. Each column identifies a different case: $k_{0}=0$, $k_{0}=10$, and $k_{0}=20$. In panels (a1)-(a3) and (b1)-(b3), the gray stripes represent the limits ($k_{1} \to 0$ and $\Gamma \to 0$) where the 2D Poisson and GinUE statistics do not hold. In panels (a1)-(a3) and (b1)-(b3), the horizontal blue (red) dashed line represents the limit of a quantum system described by 2D Poisson (GinUE) statistics. In panels (a1)-(d3), we use a system size $j=80$ that provides 12961 eigenvalues with positive parity. [(e1)-(e3)] Fraction of chaotic initial conditions in Eq.~\eqref{eq:ChaoticAttractorFraction} and [(f1)-(f3)] average Lyapunov dimension in Eq.~\eqref{eq:LyapunovDimension} as a function of the kick and dissipation strengths ($k_{1},\Gamma$). Panels (g1)-(g3) and (h1)-(h3) show the last quantities for specific dissipation strengths, $\Gamma=0.1,0.2,0.3,0.4$. Each column identifies a different case: $k_{0}=0$, $k_{0}=10$, and $k_{0}=20$. In panels (g1)-(g3), the horizontal blue (red) dashed line represents the limit of a classical system with simple (chaotic) attractors, while in panels (f1)-(f3), corresponds to the Lyapunov dimension of a point (surface). The average in phase space takes 1245 initial conditions evolved until $10^3$ periods. We use $p=2$ in all panels.}
\label{fig:QuantumClassical}
\end{figure*}

In Ref.~\cite{Grobe1988}, an ensemble average for the parameter $k_{0}$ was considered to minimize quantum fluctuations caused by the small system size examined. In this analysis, we avoid the ensemble average and instead reach the semiclassical limit by increasing the system size $j$. Additionally, we consider different values of the parameter $k_{0}$ to analyze its effect on the system. In Figs.~\ref{fig:SystemSize}(a1)-\ref{fig:SystemSize}(a3), we show the average $\langle r \rangle$ as a function of the kick strength $k_{1}$ for three cases: $k_{0}=0$, $k_{0}=10$, and $k_{0}=20$. Figures~\ref{fig:SystemSize}(b1)-\ref{fig:SystemSize}(b3) show similar plots for the average $-\langle \cos\theta \rangle$. In Figs.~\ref{fig:SystemSize}(c1)-\ref{fig:SystemSize}(c3) and Figs.~\ref{fig:SystemSize}(d1)-\ref{fig:SystemSize}(d3), we present the corresponding averages as a function of the parameter $k_{0}$ for three cases: $k_{1}=10^{-3}$, $k_{1}=1.5$, and $k_{1}=8$. In all cases, we show the averages for different system sizes. We identify how the numerical results deviate from the 2D Poisson and GinUE statistics for small system sizes. By increasing the system size, the numerical results converge to the expected statistics.

Now, we choose the maximum system size $j=80$ and present in Figs.~\ref{fig:QuantumClassical}(a1)-\ref{fig:QuantumClassical}(a3) density plots of the average $\langle r \rangle$ as a function of the kick and dissipation strengths ($k_{1},\Gamma$) for three cases: $k_{0}=0$, $k_{0}=10$, and $k_{0}=20$. Figures~\ref{fig:QuantumClassical}(b1)-\ref{fig:QuantumClassical}(b3) present similar plots for the average $-\langle \cos\theta \rangle$. Figures~\ref{fig:QuantumClassical}(c1)-\ref{fig:QuantumClassical}(c3) and Figs.~\ref{fig:QuantumClassical}(d1)-\ref{fig:QuantumClassical}(d3) show the last averages for selected values of $\Gamma$. For $k_{0}=0$, we can see an unexpected behavior for weak kick regions and low dissipation, where the system captures GinUE statistics around $k_{1}\approx0$, tends to reach 2D Poisson statistics at $k_{1}\approx1.5$, and recovers GinUE statistics for stronger kick regions. In contrast, when dissipation increases, the system tends to reach GinUE statistics for all kick strengths. The cases with $k_{0}=10$ and $k_{0}=20$ show an overall transition from weak kick regions described by 2D Poisson statistics to strong kick regions that capture GinUE statistics. Surprisingly, the last transition is irrespective of the dissipation strength $\Gamma$, due to the inclusion of the parameter $k_{0}$.

{\it Classical analysis of the kicked top.}
The classical analysis of the kicked top with dissipation, as presented in Ref.~\cite{Grobe1988}, does not consider the effects of phase space contraction induced by dissipation or those related to the phase space average of multiple initial conditions. Studying the transition to chaos in classical dynamical systems requires a well-defined invariant measure to quantify the fraction of chaos in phase space~\cite{OttBook}. In conservative systems this role is played by the phase space volume, whereas in non-conservative systems identifying such a measure is far more difficult. For leaking or absorptive systems, where excluded coordinates break phase space preservation, conditionally invariant measures can restore normalization \cite{Demers2006,Altmann2013,Altmann2013PRL}. In contrast, for dissipation induced by channels in the Lindblad master equation, which contract phase space, a practical formulation based on such invariant measures is unclear and still lacking. In view of this difficulty, and since the phase space volume measure can correctly capture both the fully regular and fully chaotic regimes, we define a fraction of chaos in direct analogy with the conservative case
 \begin{equation}
     \label{eq:ChaoticAttractorFraction}
     f_{c} = \frac{1}{V_{0}}\int_{\mathcal{M}} dQ\,dP \, \Upsilon(Q,P),
 \end{equation} 
where $\Upsilon$ is a characteristic function that takes the value $\Upsilon=1$ if the maximal Lyapunov exponent of an evolved initial condition is positive, or $\Upsilon=0$ otherwise~\cite{foot1}. The classical variables $J_{x,y,z}$ can be transformed to the variables $Q$ and $P$, which provide a bounded phase space $\mathcal{M} = \{(Q,P)|Q^{2}+P^{2} \leq 4\}$ with volume $V_{0}=\int_{\mathcal{M}} dQ\,dP = 4\pi$ (see the Supplemental Material~\cite{footSM} for further details). We utilize the phase space $\mathcal{M}$ linked to the initial conditions to average the characteristic function $\Upsilon$ at time $t=0$ before the phase space contraction, scaling the average by the initial phase space volume $V_{0}$. In this way, $f_{c}=1$ defines a classical system completely governed by chaotic attractors, and $f_{c}=0$ a system with only simple attractors. Additionally, we use the Lyapunov dimension~\cite{Kaplan1979,Russell1980,Mori1980,Farmer1983}
\begin{equation}
    \label{eq:LyapunovDimension}
    D_{\text{L}} = k + \frac{1}{|h_{k+1}|}\sum_{i=1}^{k}h_{i},
\end{equation}
where $k<N$ is the largest index of the Lyapunov spectrum of an $N$-dimensional system, $\{h_{1}>\ldots>h_{N}\}$, which fulfills the constriction $\sum_{i=1}^{k}h_{i} > 0$. The Lyapunov dimension constitutes an upper limit to the Hausdorff or fractal dimension~\cite{MandelbrotBook,OttBook}, which provides a measure of a chaotic attractor. In the Supplemental Material~\cite{footSM}, we present the correspondence between the Lyapunov dimension and the Hausdorff dimension.

We use the same parameters from the quantum analysis of the kicked top with dissipation to conduct its classical analysis. In Figs.~\ref{fig:QuantumClassical}(e1)-\ref{fig:QuantumClassical}(e3), we show density plots of the fraction $f_{c}$ in Eq.~\eqref{eq:ChaoticAttractorFraction} as a function of the parameters $k_{1}$ and $\Gamma$, for the previous cases: $k_{0}=0$, $k_{0}=10$, and $k_{0}=20$. For $k_{0}=0$, we identify a wide region of regular motion for weak kick regions, where point attractors or limit cycles appear. In contrast, the motion is dominated by chaotic attractors for stronger kick regions. The cases with $k_{0}=10$ and $k_{0}=20$ show a reduction of the regular region for weak kick regions and low dissipation. In these cases, the classical transition depends on the dissipation strength, as can be seen in Figs.~\ref{fig:QuantumClassical}(g1)-\ref{fig:QuantumClassical}(g3) for selected values of $\Gamma$, where the highest dissipation produces a delayed transition to the chaotic regime. Complementarily, Figs.~\ref{fig:QuantumClassical}(f1)-\ref{fig:QuantumClassical}(f3) show density plots of the average Lyapunov dimension $\overline{D}_{\text{L}}$ in Eq.~\eqref{eq:LyapunovDimension} for the presented cases. Figures~\ref{fig:QuantumClassical}(h1)-\ref{fig:QuantumClassical}(h3) show $\overline{D}_{\text{L}}$ for some values of $\Gamma$. Regular regions dominated by point attractors show a null Lyapunov dimension, which suddenly changes to a unit value in regions with limit cycles. After this, the Lyapunov dimension increases and reaches values slightly below $\overline{D}_{\text{L}}=2$, where the regions contain chaotic attractors.

{\it Correspondence principle in the kicked top.}
There is an alignment between the quantum and classical perspectives in the strong kick regions. However, this correspondence breaks down in the weak kick regions combined with high dissipation. The disagreement is more evident for the case with $k_{0}=0$. The cases with $k_{0}\neq0$ lead to a quantum transition that is independent of the dissipation strength. This leads to partial agreement within the weak kick regions when dissipation is very low ($\Gamma\leq0.1$) and the dissipative system tends to behave like the isolated one (see the Supplemental Material~\cite{footSM} for a comparison). However, for higher dissipation, the classical transition identified in the cases with $k_{0}=10$ and $k_{0}=20$ clearly depends on the dissipation strength, in contrast to the quantum transition.

As reported in Ref.~\cite{Grobe1988}, for the parameters $k_{0}=10$ and $\Gamma=0.1$ shown in Fig.~\ref{fig:QuantumClassical}, the individual case with $k_{1}=0$ shows an agreement between classical regular motion and 2D Poisson statistics. In contrast, the case with $k_{1}=8$ demonstrates alignment between classical chaotic motion and GinUE statistics. However, for higher dissipation ($0.1<\Gamma\leq0.4$), the quantum-classical correspondence gradually breaks down in the weak kick region. There is a clear region in Fig.~\ref{fig:QuantumClassical}(c2) and Fig.~\ref{fig:QuantumClassical}(d2), $0\leq k_{1}<2$, where the statistics developed by simple attractors [see Fig.~\ref{fig:QuantumClassical}(g2) and Fig.~\ref{fig:QuantumClassical}(h2)] deviate from 2D Poisson statistics and tend to reach GinUE statistics. A similar disagreement is identified when comparing Figs.~\ref{fig:QuantumClassical}(c3)-\ref{fig:QuantumClassical}(d3) with Figs.~\ref{fig:QuantumClassical}(g3)-\ref{fig:QuantumClassical}(h3) for the case with $k_{0}=20$. The mechanism of why a simple attractor could develop GinUE spectral correlations is still an open question~\cite{Villasenor2024}.

{\it Discussion.}
The study presented in this work demonstrated that the correspondence principle fails in the kicked system that inspired the current definition of dissipative quantum chaos. This is supported by achieving a genuine semiclassical limit for the quantum system. As a result, Ginibre spectral correlations cannot be used as a test of dissipative quantum chaos. This observation encourages new approaches to explore this phenomenon. Important remarks arise in this regard. (1) The concept of dissipative quantum chaos should be grounded in a robust framework that incorporates a clear classical interpretation and is not solely based on spectral correlations. (2) Identifying the quantum fingerprints of classical chaotic structures appears to be a promising way to address this problem~\cite{Dutta2025,Cai2025}.

{\it Acknowledgments---}
We acknowledge the help of the supercomputer system HPC Vega - IZUM under project No. S24O02-01. This work was supported by the Slovenian Research and Innovation Agency (ARIS) under Grants No. J1-4387 and No. P1-0306.

\bibliography{main}

\begin{thebibliography}{145}
\bibitem [{foo(The maximal Lyapunov exponent with zero value identifies a limit cycle, while a negative value is associated to a point attractor)}]{foot1}%
  \BibitemOpen
  \href@noop {} {} (\bibinfo {year} {The maximal Lyapunov exponent with zero value identifies a limit cycle, while a negative value is associated to a point attractor})\BibitemShut {NoStop}%
\end{thebibliography}

\newpage
\cleardoublepage

\begin{widetext}

\setcounter{equation}{0}
\setcounter{figure}{0}

\makeatletter 
\renewcommand{\thesection}{S\@Roman\c@section}
\renewcommand{\thefigure}{S\@arabic\c@figure}
\renewcommand{\theequation}{S\@arabic\c@equation}
\makeatother

\begin{center}
    {\large \textbf{Supplemental Material: Correspondence principle, dissipation, and Ginibre ensemble} \\}
    \vspace{0.2in}
    David Villase\~nor$^{1}$, Hua Yan$^{1}$, Matic Orel$^{1}$, Marko Robnik$^{1}$ \\
    \vspace{0.1in}
    {\small \it{$^{1}$CAMTP - Center for Applied Mathematics and Theoretical Physics, \\ University of Maribor, Mladinska 3, SI-2000 Maribor, Slovenia, European Union} \\}
\end{center}

This Supplemental Material provides additional details on the kicked top with dissipation. In Sec.~\ref{sec:IsolatedKickedTop}, we discuss both classical and quantum aspects of the kicked top in the absence of dissipation. In Sec.~\ref{sec:ComplexSpectrum}, we analyze the characteristics of the kicked top with dissipation, focusing on the numerical randomness of the complex spectrum that arises under high dissipation and the correspondence principle for low dissipation. Finally, in Sec.~\ref{sec:DissipationProcess}, we analyze an approximate description of the dissipation process and compare it with the description presented in the main text.

\section{Kicked top in absence of dissipation}
\label{sec:IsolatedKickedTop}

\subsection{Heisenberg equations and map of isolated classical motion}

The Heisenberg equations provide the quantum evolution of the angular momentum operators $\hat{J}_{x,y,z}$. This evolution can be identified by a set of rotations
\begin{equation}
    \hat{J}_{x,y,z}^{n+1} = \hat{F}^{\dag}\hat{J}_{x,y,z}^{n}\hat{F} = \hat{R}(\hat{\Omega}_{1}^{n})\hat{R}(\hat{\Omega}_{0}^{n})\hat{R}(p)\hat{J}_{x,y,z}^{n} ,
\end{equation}
where $\hat{F}=e^{-i\hat{H}_{1}}e^{-i\hat{H}_{0}}$ is the Floquet operator of the kicked top in absence of dissipation, $\hat{H}_{0} = p\hat{J}_{z} + \frac{k_{0}}{2j}\hat{J}_{z}^{2}$, and $\hat{H}_{1} = \frac{k_{1}}{2j}\hat{J}_{y}^{2}$. The rotation matrices are given by
\begin{equation}
    \hat{R}(p) = \left(\begin{array}{ccc}
         \cos(p) & -\sin(p) & 0 \\
         \sin(p) & \cos(p) & 0 \\
         0 & 0 & 1 \end{array}\right) , \,
    \hat{R}(\hat{\Omega}_{0}^{n}) = \left(\begin{array}{ccc}
         \cos(\hat{\Omega}_{0}^{n}) & -\sin(\hat{\Omega}_{0}^{n}) & 0 \\
         \sin(\hat{\Omega}_{0}^{n}) & \cos(\hat{\Omega}_{0}^{n}) & 0 \\
         0 & 0 & 1 \end{array}\right) , \,
    \hat{R}(\hat{\Omega}_{1}^{n}) = \left(\begin{array}{ccc}
         \cos(\hat{\Omega}_{1}^{n}) & 0 & \sin(\hat{\Omega}_{1}^{n}) \\
         0 & 1 & 0 \\
         -\sin(\hat{\Omega}_{1}^{n}) & 0 & \cos(\hat{\Omega}_{1}^{n}) 
          \end{array}\right) 
\end{equation}
with rotation angles
\begin{equation}
    \hat{\Omega}_{0}^{n} = \frac{k_{0}}{2j}\left(2\hat{J}_{z}^{n}+1\right) , \hspace{0.5in} \hat{\Omega}_{1}^{n} = \frac{k_{1}}{2j}\left[2\left(\sin(\hat{\Omega}_{0}^{n})\hat{R}_{x}^{n}(p)+\cos(\hat{\Omega}_{0}^{n})\hat{R}_{y}^{n}(p)\right)+1\right] ,
\end{equation}
where $\hat{R}_{x}^{n}(p) = \cos(p)\hat{J}_{x}^{n}-\sin(p)\hat{J}_{y}^{n}$ and $\hat{R}_{y}^{n}(p) = \sin(p)\hat{J}_{x}^{n}+\cos(p)\hat{J}_{y}^{n}$.

The mean-field approximation of the last expressions using Bloch coherent states, $|\alpha\rangle = e^{(\alpha\hat{J}_{+}-\alpha^{\ast}\hat{J}_{-})}|j,-j\rangle$, of minimum uncertainty provides the map which describes the isolated classical motion of the kicked top
\begin{equation}
    \label{eq:IsolatedClassicalMap}
    J_{x,y,z}^{n+1} = \lim_{j\to\infty}\frac{1}{j}\langle \alpha|\hat{J}_{x,y,z}^{n+1}|\alpha\rangle = R(\Omega_{1}^{n})R(\Omega_{0}^{n})R(p)J_{x,y,z}^{n} ,
\end{equation}
where we identify the expressions
\begin{equation}
    R(p) = \left(\begin{array}{ccc}
         \cos(p) & -\sin(p) & 0 \\
         \sin(p) & \cos(p) & 0 \\
         0 & 0 & 1 \end{array}\right) , \,
    R(\Omega_{0}^{n}) = \left(\begin{array}{ccc}
         \cos(\Omega_{0}^{n}) & -\sin(\Omega_{0}^{n}) & 0 \\
         \sin(\Omega_{0}^{n}) & \cos(\Omega_{0}^{n}) & 0 \\
         0 & 0 & 1 \end{array}\right) , \,
    R(\Omega_{1}^{n}) = \left(\begin{array}{ccc}
         \cos(\Omega_{1}^{n}) & 0 & \sin(\Omega_{1}^{n}) \\
         0 & 1 & 0 \\
         -\sin(\Omega_{1}^{n}) & 0 & \cos(\Omega_{1}^{n})
          \end{array}\right) ,
\end{equation}
with
\begin{equation}
    \Omega_{0}^{n} = k_{0}J_{z}^{n} , \hspace{0.5in} \Omega_{1}^{n} = k_{1}\left(\sin(\Omega_{0}^{n})R_{x}^{n}(p)+\cos(\Omega_{0}^{n})R_{y}^{n}(p)\right) ,
\end{equation}
$R_{x}^{n}(p) = \cos(p)J_{x}^{n}-\sin(p)J_{y}^{n} $, and $R_{y}^{n}(p) = \sin(p)J_{x}^{n}+\cos(p)J_{y}^{n}$.

\subsection{Stereographic projection of the Bloch sphere}

The natural pair of conjugated canonical variables on the Bloch sphere consists of the azimuthal angle $\phi$ and the cosine of the polar angle $\cos\theta$, which have a Poisson bracket given by $\{\phi,\cos\theta\}=1$. These variables correspond to the angular momentum components $J_{x,y,z}$ through the relationships $\phi=\arctan(J_{y}/J_{x})$ and $\cos\theta=J_{z}$. However, the angular momentum variables are non-canonical, with the Poisson bracket $\{J_{x},J_{y}\}=J_{z}$.

By using a canonical transformation defined as $Q=\sqrt{2(1+\cos\theta)}\cos\phi$ and $P=-\sqrt{2(1+\cos\theta)}\sin\phi$, we establish a Poisson bracket $\{Q,P\}=1$. This transformation allows the angular momentum variables on the Bloch sphere to be mapped onto conjugate position and momentum variables in a plane
\begin{align}
    \left(\begin{array}{c}
       Q \\
       P
    \end{array}\right) = 
    \left(\begin{array}{c}
       J_{x} \sqrt{\frac{2 (1 + J_{z})}{J_{x}^{2} + J_{y}^{2}}} \\
       -J_{y} \sqrt{\frac{2 (1 + J_{z})}{J_{x}^{2} + J_{y}^{2}}}
    \end{array}\right) \leftrightarrow
    \left(\begin{array}{c}
       J_{x} \\
       J_{y} \\
       J_{z}
    \end{array}\right) = 
    \left(\begin{array}{c}
       Q \sqrt{1 - \frac{Q^{2}+P^{2}}{4}} \\
       -P \sqrt{1 - \frac{Q^{2}+P^{2}}{4}} \\
       \frac{Q^{2}+P^{2}}{2} - 1
    \end{array}\right) .
\end{align}
The last transformation provides a bounded phase space $\mathcal{M} = \{(Q,P)|Q^{2}+P^{2} \leq 4\}$, which avoids the periodicity of the phase space given by the angular variables $(\phi,\cos\theta)$.

\subsection{Correspondence principle in absence of dissipation}

We investigate the correspondence principle between classical and quantum signatures of chaos for the kicked top in absence of dissipation using the system parameters presented in Ref.~\cite{Grobe1988}. For the classical analysis, we calculate the fraction of chaotic regions within the available phase space using the quantity
\begin{equation}
    \label{eq:ChaosFraction}
    \mu_{c} = \frac{1}{V_{0}}\int_{\mathcal{M}} dQ\,dP \, \chi(Q,P) ,
\end{equation}
where $\chi$ is a characteristic function that depends on the coordinates $(Q,P)$ and takes the values 0 or 1. We set $\chi=1$ if the Lyapunov exponent of an evolved initial condition is positive, otherwise $\chi=0$. In this way, $\mu_{c}=1$ defines a fully chaotic classical system and $\mu_{c}=0$ an integrable one. The volume of the available phase space is given by $V_{0}=\int_{\mathcal{M}} dQ\,dP = 4\pi$.

In our quantum analysis, we utilize the real eigenphases $\phi_{k}$ derived from the complex eigenvalues of the Floquet operator associated with the kicked top. These eigenvalues (eigenphases) satisfy the eigenvalue equation $\hat{F}|\nu_{k}\rangle = \nu_{k}|\nu_{k}\rangle = e^{i\phi_{k}}|\nu_{k}\rangle$. The spectral analysis of the spacings between consecutive eigenvalues, $s_{k} = \phi_{k+1}-\phi_{k}$, reveals that for integrable systems, the spacing distribution follows a Poisson distribution~\cite{Berry1977}, expressed as $P(s)=e^{-s}$. In contrast, chaotic systems exhibit a spacing distribution described by the Wigner-Dyson ensemble~\cite{Bohigas1984}, given by $P(s)=\frac{\pi}{2}e^{-\pi s^{2}/4}$, which corresponds to the distribution of the Gaussian orthogonal ensemble (GOE) of random matrices for conservative systems, or equivalently, the circular orthogonal ensemble (COE)~\cite{MehtaBook} for periodically kicked systems. To eliminate the need for the unfolding procedure~\cite{HaakeBook,StockmannBook,Guhr1998}, we employ the spectral ratio~\cite{Oganesyan2007,Atas2013}
\begin{equation}
    r_{k} = \frac{\min(\phi_{k+1}-\phi_{k},\phi_{k}-\phi_{k-1})}{\max(\phi_{k+1}-\phi_{k},\phi_{k}-\phi_{k-1})} ,
\end{equation}
which provides the mean values $\langle r\rangle_{\text{P}} \approx 0.386$ and $\langle r\rangle_{\text{COE}} \approx 0.536$ for the Poisson and COE distributions, respectively. Using the last mean values, we can define a normalized spectral ratio
\begin{equation}
    \label{eq:SpectralRatio}
    r_{c} = \frac{\langle r\rangle - \langle r\rangle_{\text{P}}}{\langle r\rangle_{\text{COE}} - \langle r\rangle_{\text{P}}} ,
\end{equation}
where $r_{c}=1$ identifies a fully chaotic quantum system, while $r_{c}=0$ represents an integrable quantum system. 

In Fig.~\ref{fig:QuantumClassicalIsolated}, we illustrate the relationship between the classical and quantum signatures of chaos for the kicked top in the absence of dissipation, as a function of the kick strength. We utilize the system parameters outlined in Ref.~\cite{Grobe1988} and investigate the effect of the parameter $k_{0}$ on the system behavior. For the classical analysis, we evolve the equations of motion presented in Eq.~\eqref{eq:IsolatedClassicalMap} for a set of initial conditions and compute the maximal Lyapunov exponent. We then calculate the chaos fraction using Eq.~\eqref{eq:ChaosFraction} and present the results in Fig.~\ref{fig:QuantumClassicalIsolated}(a1) for $k_{0}=0$, Fig.~\ref{fig:QuantumClassicalIsolated}(a2) for $k_{0}=10$, and Fig.~\ref{fig:QuantumClassicalIsolated}(a3) for $k_{0}=20$. For the quantum analysis, we diagonalize the Floquet operator of the kicked top and utilize two sectors of eigenvalues (eigenphases) with positive and negative parity to compute the normalized spectral ratio as defined in Eq.~\eqref{eq:SpectralRatio}. The results are presented in Fig.~\ref{fig:QuantumClassicalIsolated}(b1) for $k_{0}=0$, Fig.~\ref{fig:QuantumClassicalIsolated}(b2) for $k_{0}=10$, and Fig.~\ref{fig:QuantumClassicalIsolated}(b3) for $k_{0}=20$.

\begin{figure}[ht]
\centering
\includegraphics[width=0.95\textwidth]{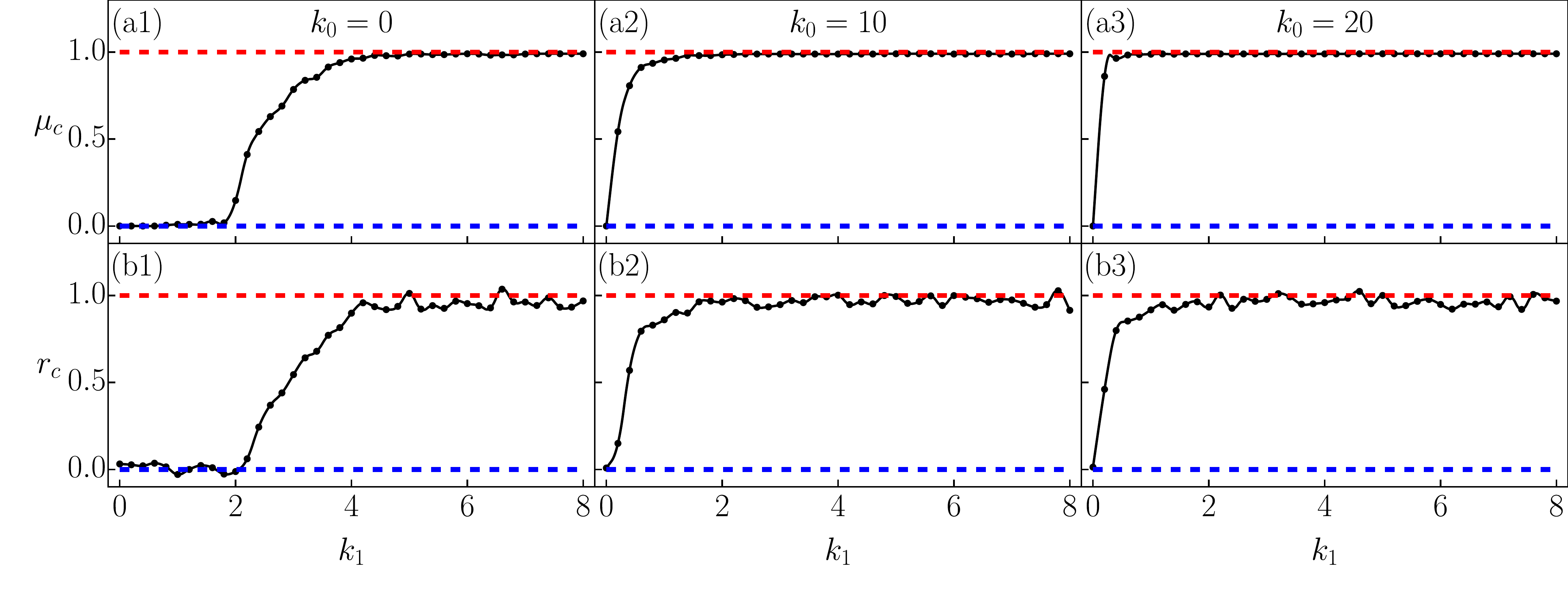}
\caption{Classical and quantum signatures of chaos for the kicked top in absence of dissipation. [(a1)-(a3)] Chaos fraction in Eq.~\eqref{eq:ChaosFraction} and [(b1)-(b3)] normalized spectral ratio in Eq.~\eqref{eq:SpectralRatio} as a function of the kick strength $k_{1}$. Each column represents a different case: [(a1)-(b1)] $k_{0}=0$, [(a2)-(b2)] $k_{0}=10$, and [(a3)-(b3)] $k_{0}=20$. In panels (a1)-(a3), we evolve a set of 1250 initial conditions until $10^{3}$ periods for each kick strength. The horizontal blue (red) dashed line represents the limit for a regular (chaotic) classical system. In panels (b1)-(b3), we chose a system size $j=4096$ and average the spectral ratios from the two parity sectors of the Floquet operator for each kick strength. The horizontal blue (red) dashed line represents the limit for a regular (chaotic) quantum system. System parameters: $p=2$ and $\Gamma=0$.}
\label{fig:QuantumClassicalIsolated}
\end{figure}

In the previous figures, we observe a strong correlation between the classical and quantum signatures of chaos for all values of $k_{0}$. Regular classical regions correspond to Poisson statistics, while chaotic classical regions align with COE statistics. See the columns: Figs.~\ref{fig:QuantumClassicalIsolated}(a1)-\ref{fig:QuantumClassicalIsolated}(b1) for $k_{0}=0$, Figs.~\ref{fig:QuantumClassicalIsolated}(a2)-\ref{fig:QuantumClassicalIsolated}(b2) for $k_{0}=10$, and Figs.~\ref{fig:QuantumClassicalIsolated}(a3)-\ref{fig:QuantumClassicalIsolated}(b3) for $k_{0}=20$. Furthermore, we find that varying the parameter $k_{0}$ significantly influences the transition from regularity to chaos. For $k_{0}=0$, this transition occurs around $k_{1}\approx2$ [see Figs.~\ref{fig:QuantumClassicalIsolated}(a1)-\ref{fig:QuantumClassicalIsolated}(b1), column view]. In the case of $k_{0}=10$, the transition occurs rapidly [see Figs.~\ref{fig:QuantumClassicalIsolated}(a2)-\ref{fig:QuantumClassicalIsolated}(b2), column view]. Lastly, for $k_{0}=20$, the transition is almost immediate [see Figs.~\ref{fig:QuantumClassicalIsolated}(a3)-\ref{fig:QuantumClassicalIsolated}(b3), column view]. This finding emphasizes the important role of the parameter $k_{0}$ in the classical dynamics of the isolated system.

\section{Kicked top with dissipation}
\label{sec:ComplexSpectrum}

\subsection{Numerical randomness in the complex spectrum of the kicked top}

The numerical eigenvalues of the dissipative Floquet operator for the kicked top collapse towards the center of the unit circle at high dissipation strengths~\cite{Braun1999PhysD}. Consequently, these eigenvalues fall below the machine precision threshold, $\epsilon=10^{-16}$, and can be considered essentially random numbers. In Fig.~\ref{fig:SpectrumCollapse}, we illustrate the collapse of the numerical spectrum as dissipation increases for the previous cases of $k_{0}=0,10,20$. We examine two values of kick strength, one weak ($k_{1}=10^{-3}$) and one strong ($k_{1}=8$). Specifically, Fig.~\ref{fig:SpectrumCollapse}(a4), Fig.~\ref{fig:SpectrumCollapse}(b4), and Fig.~\ref{fig:SpectrumCollapse}(c4) display the spectrum collapse to the origin of the unit circle under the strongest dissipation conditions with a weak kick. In contrast, Fig.~\ref{fig:SpectrumCollapse}(d4), Fig.~\ref{fig:SpectrumCollapse}(e4), and Fig.~\ref{fig:SpectrumCollapse}(f4) illustrate the same collapse occurring with a strong kick.

\begin{figure}[ht]
\centering
\includegraphics[width=0.95\textwidth]{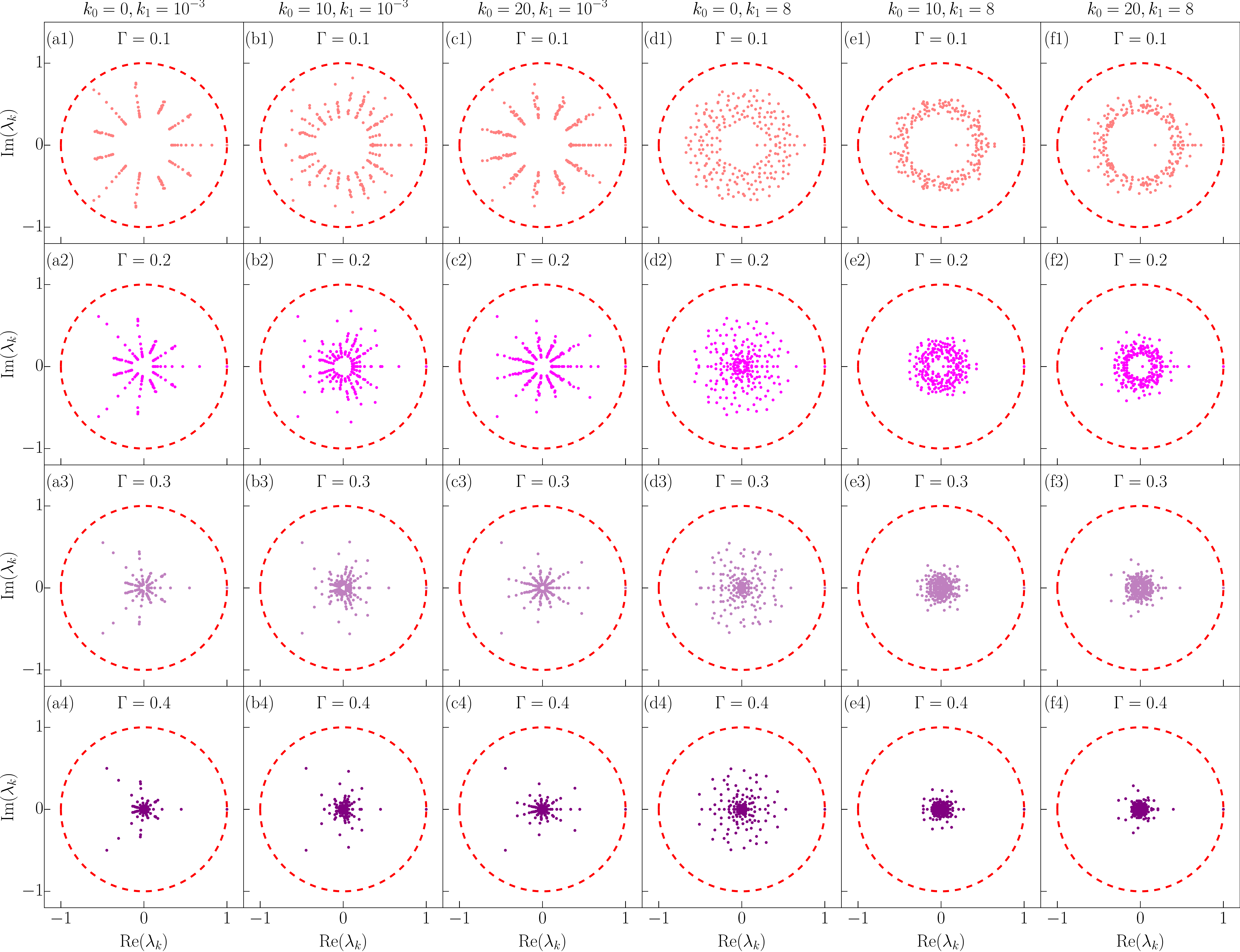}
\caption{Complex eigenvalues of the dissipative Floquet operator of the kicked top. The first three columns identify a weak kick $k_{1}=10^{-3}$: [(a1)-(a4)] $k_{0}=0$, [(b1)-(b4)] $k_{0}=10$, and [(c1)-(c4)] $k_{0}=20$. The last three columns identify a strong kick $k_{1}=8$: [(d1)-(d4)] $k_{0}=0$, [(e1)-(e4)] $k_{0}=10$, and [(f1)-(f4)] $k_{0}=20$. In contrast, each row identifies a different dissipation strength: [(a1)-(f1)] $\Gamma=0.1$, [(a2)-(f2)] $\Gamma=0.2$, [(a3)-(f3)] $\Gamma=0.3$, and [(a4)-(f4)] $\Gamma=0.4$. System parameters: $p=2$ and $j=10$.}
\label{fig:SpectrumCollapse}
\end{figure}

\begin{figure}[ht]
\centering
\includegraphics[width=0.95\textwidth]{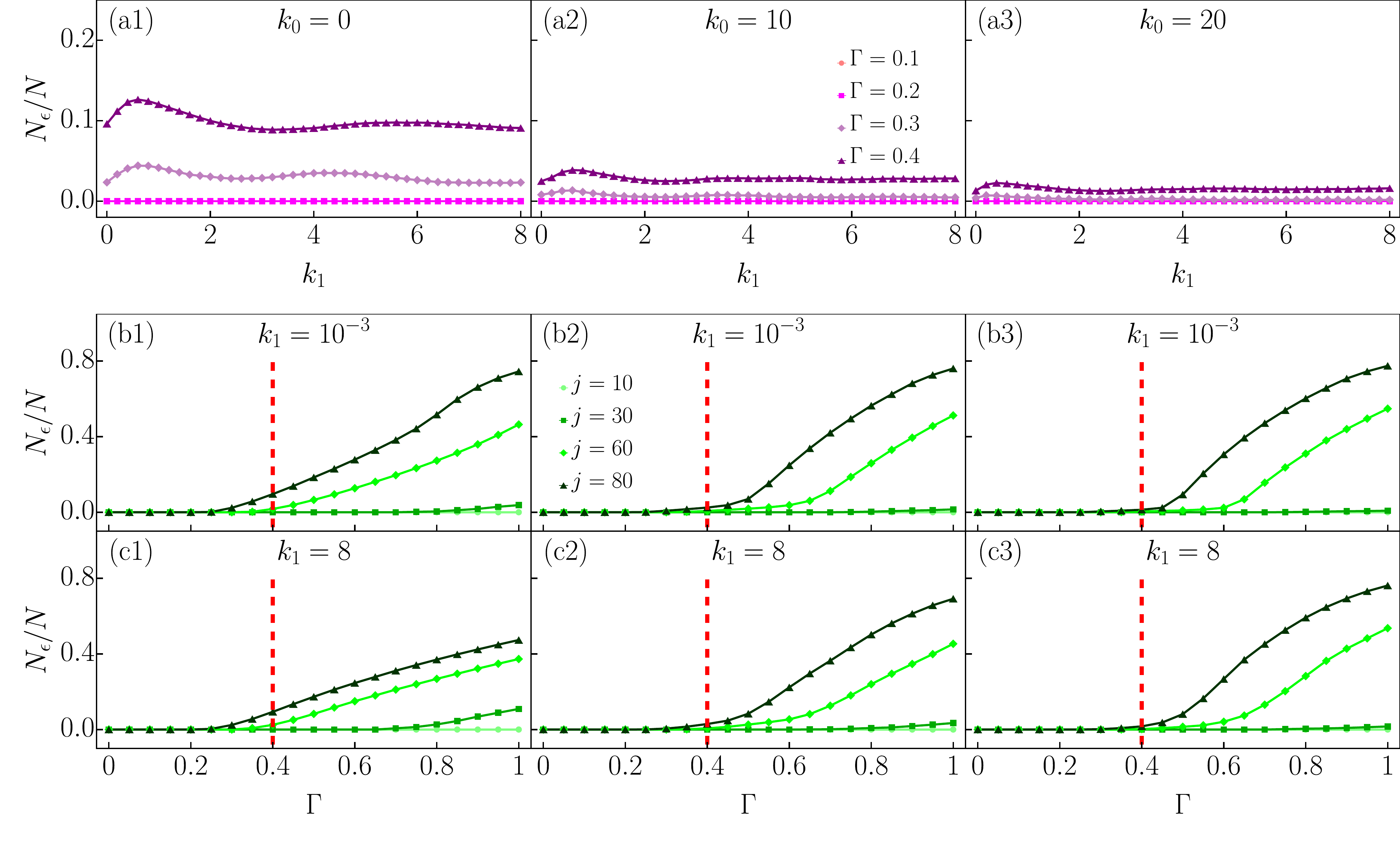}
\caption{[(a1)-(a3)] Fraction of eigenvalues of the dissipative Floquet operator, which are below the machine precision, as a function the kick strength $k_{1}$. Fraction of eigenvalues below the machine precision as a function of the dissipation strength $\Gamma$ for [(b1)-(b3)] weak kick $k_{1}=10^{-3}$ and [(c1)-(c3)] strong kick $k_{1}=8$. Each column represents a different case: [(a1)-(c1)] $k_{0}=0$, [(a2)-(c2)] $k_{0}=10$, and [(a3)-(c3)] $k_{0}=20$. In panels (a1)-(a3), we chose different dissipation strengths $\Gamma=0.1,0.2,0.3,0.4$ and the system size $j=80$. In panels (b1)-(b3) and (c1)-(c3), we chose different system sizes $j=10,30,60,80$ and the vertical red dashed line represents the dissipation threshold at $\Gamma=0.4$. We use $p=2$ in all panels.}
\label{fig:ErrorAnalysis}
\end{figure}

In the study described in the main text, we limited the dissipation strength to the interval $0<\Gamma\leq0.4$ based on numerical tests. For $\Gamma>0.4$, the number of eigenvalues below the machine precision, denoted as $N_{\epsilon}$, is comparable to the total number of eigenvalues $N$ obtained from the numerical diagonalization of the dissipative Floquet operator. In Fig.~\ref{fig:ErrorAnalysis}, we provide a detailed account of these numerical tests, where we computed the eigenvalues for varying dissipation strengths and system sizes. Figure~\ref{fig:ErrorAnalysis}(a1) displays the fraction of eigenvalues $N_{\epsilon}/N$ as a function of the kick strength for $k_{0}=0$. Figure~\ref{fig:ErrorAnalysis}(a2) and Fig.~\ref{fig:ErrorAnalysis}(a3) show the same fraction for $k_{0}=10$ and $k_{0}=20$, respectively. We calculated the fraction $N_{\epsilon}/N$ using different dissipation strengths while maintaining a fixed system size of $j=80$. Our findings indicate that the behavior of the fraction is not significantly affected by the kick strength. Notably, for the case of $k_{0}=0$, the fraction $N_{\epsilon}/N$ increases at a faster rate compared to the cases where $k_{0}\neq0$.

In Fig.~\ref{fig:ErrorAnalysis}(b1), we present the fraction of eigenvalues $N_{\epsilon}/N$ as a function of the dissipation strength for $k_{0}=0$ and a weak kick ($k_{1}=10^{-3}$). The same fraction is shown in Fig.~\ref{fig:ErrorAnalysis}(b2) and Fig.~\ref{fig:ErrorAnalysis}(b3) for $k_{0}=10$ and $k_{0}=20$, respectively. Additionally, we compute the fraction \(N_{\epsilon}/N\) for a stronger kick ($k_{1}=8$), with the corresponding results presented in Figs.~\ref{fig:ErrorAnalysis}(c1)-\ref{fig:ErrorAnalysis}(c3). Our analysis indicates that the fraction $N_{\epsilon}/N$ begins to increase with high dissipation and larger system sizes. In contrast, for smaller system sizes, the growth is less significant, even under high dissipation. From our numerical tests, we identify a range of dissipation strength where the fraction eigenvalues below the machine precision is negligible: $0<\Gamma\leq0.4$. The threshold value $\Gamma=0.4$ is marked by a vertical red dashed line in Figs.~\ref{fig:ErrorAnalysis}(b1)-\ref{fig:ErrorAnalysis}(b3) and Figs.~\ref{fig:ErrorAnalysis}(c1)-\ref{fig:ErrorAnalysis}(c3).

\newpage

\subsection{Lyapunov dimension and Hausdorff dimension}

In our study of classical dynamics in the kicked top with dissipation, we used the Lyapunov dimension because calculating the Hausdorff dimension is numerically more complex and less stable. The Lyapunov dimension $D_{\text{L}}$, as presented in Eq.~(16) of the main text, serves as an upper limit to the Hausdorff or fractal dimension~\cite{MandelbrotBook,OttBook}. In some systems, these two dimensions can be equivalent~\cite{Russell1980}, as occurs in the kicked top with dissipation. The Hausdorff dimension measures the complexity of a chaotic attractor, such as the well-known Lorenz attractor~\cite{Mori1980,Viswanath2004}, and is defined as~\cite{OttBook}
\begin{equation}
    \label{eq:Hausdorff}
    D_{\text{H}} = - \lim_{\epsilon \to 0} \frac{\ln C(\epsilon)}{\ln \epsilon} ,
\end{equation}
where $\epsilon$ is the side length of a square cell in a two-dimensional phase space and $C(\epsilon)$ counts the number of cells needed to cover the attractor subset of the phase space. For simple geometrical structures such as a point, $C(\epsilon)$ is independent of $\epsilon$ and its Hausdorff dimension is $D_{\text{H}}=0$. For other structures as a line, $C(\epsilon)\sim1/\epsilon$ and $D_{\text{H}}=1$. If the structure is a surface, $C(\epsilon)\sim1/\epsilon^{2}$ and $D_{\text{H}}=2$.

The chaotic attractors identified in the classical dynamics of the kicked top with dissipation exhibit an average Lyapunov dimension or Hausdorff dimension very close to the dimension of the isolated system, which coincides with the dimension of a surface $\overline{D}_{\text{L}}=2$. In Fig.~\ref{fig:BifurcationDiagram}, we provide a more detailed analysis of the chaotic attractors. We start at the south pole on the Bloch sphere, defined by the coordinates $(J_{x},J_{y},J_{z})=(0,0,-1)$, and evolve this point under strong kick $k_{1}=8$ with varying dissipation strengths for the cases of $k_{0}=0,10,20$. In Fig.~\ref{fig:BifurcationDiagram}(a1), we present the bifurcation diagram of the angular momentum variable $J_{y}$ as a function of the dissipation strength for $k_{0}=0$. Figure~\ref{fig:BifurcationDiagram}(a2) and Fig.~\ref{fig:BifurcationDiagram}(a3) illustrate the bifurcation diagrams for $k_{0}=10$ and $k_{0}=20$, respectively. Figures~\ref{fig:BifurcationDiagram}(b1)-\ref{fig:BifurcationDiagram}(b3) display the maximum and minimum Lyapunov exponents, $h_{\max}$ and $h_{\min}$, as functions of dissipation strength for the respective cases of $k_{0}=0,10,20$. We observe that, under high dissipation, the disordered patterns of the bifurcation diagrams merge into a single line that indicates a point attractor. The agreement aligns with the maximal Lyapunov behavior, where positive values emerge in disordered regions and negative values are found in the single line regions, as illustrated in the bifurcation diagrams.

\begin{figure}[ht]
\centering
\includegraphics[width=0.95\textwidth]{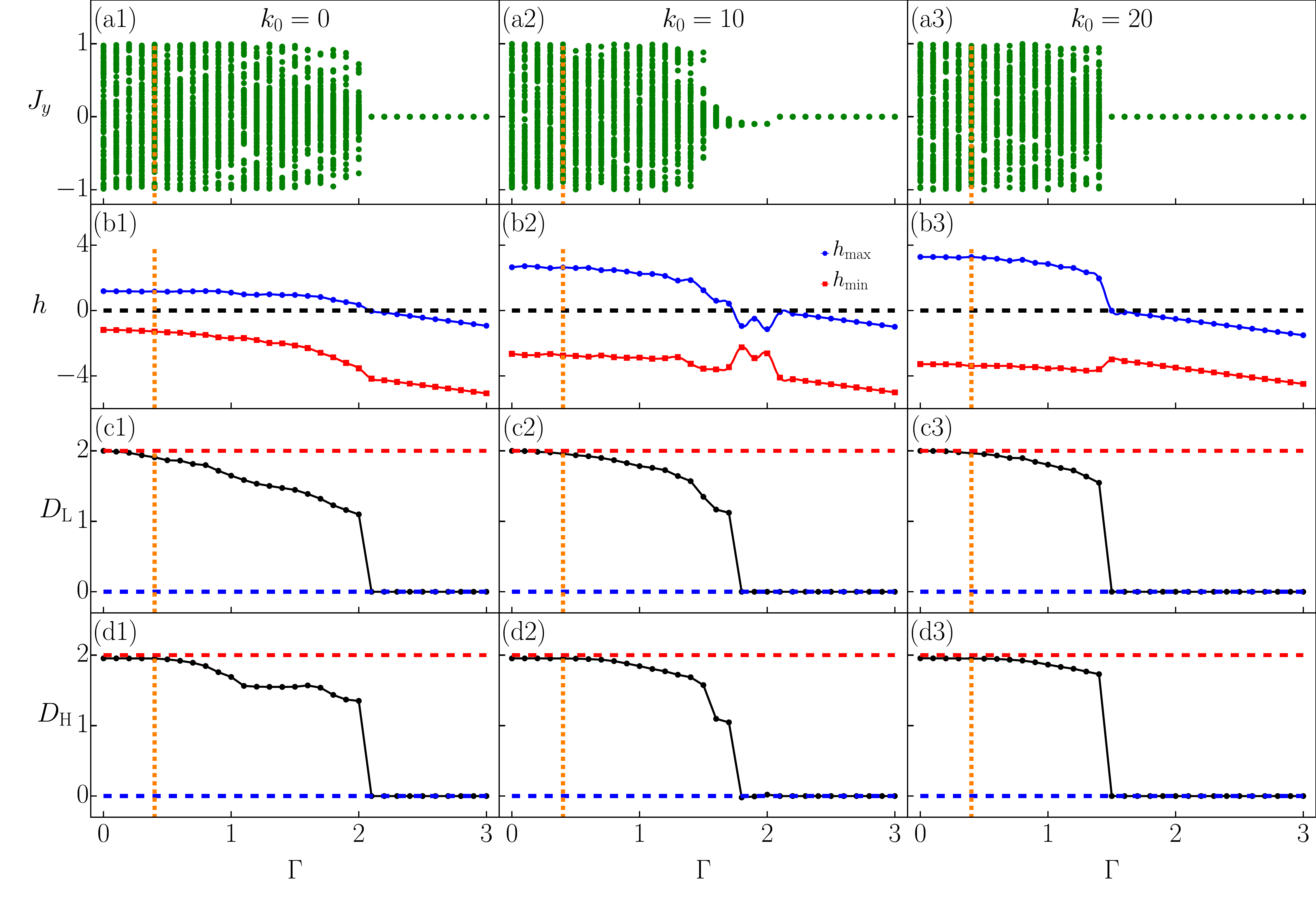}
\caption{[(a1)-(a3)] Bifurcation diagram of the angular momentum variable $J_{y}$ as a function of the dissipation strength $\Gamma$. [(b1)-(b3)] Maximal ($h_{\max}$, blue line) and minimal ($h_{\min}$, red line) Lyapunov exponent as a function of the dissipation strength. [(c1)-(c3)] Lyapunov dimension in Eq.~(16) of the main text and [(d1)-(d3)] Hausdorff dimension in Eq.~\eqref{eq:Hausdorff} as a function of the dissipation strength. Each column represents a different case: [(a1)-(d1)] $k_{0}=0$, [(a2)-(d2)] $k_{0}=10$, and [(a3)-(d3)] $k_{0}=20$. In panels (a1)-(a3), we plot the last 100 periods of the classical evolution. In panels (c1)-(c3) and (d1)-(d3), the horizontal blue (red) dashed line represents the Lyapunov dimension of a point (surface). In all panels, the vertical orange dotted line represents the dissipation threshold at $\Gamma=0.4$. Initial condition evolved until $10^{3}$ periods: $(J_{x},J_{y},J_{z})=(0,0,-1)$. The same initial condition was evolved $10^6$ periods for the calculation of Hausdorff dimension with cell side length $\epsilon\in[2^{-5.05},2^{-3}]$. System parameters: $p=2$ and $k_{1}=8$.}
\label{fig:BifurcationDiagram}
\end{figure}

\begin{figure}[ht]
\centering
\includegraphics[width=0.95\textwidth]{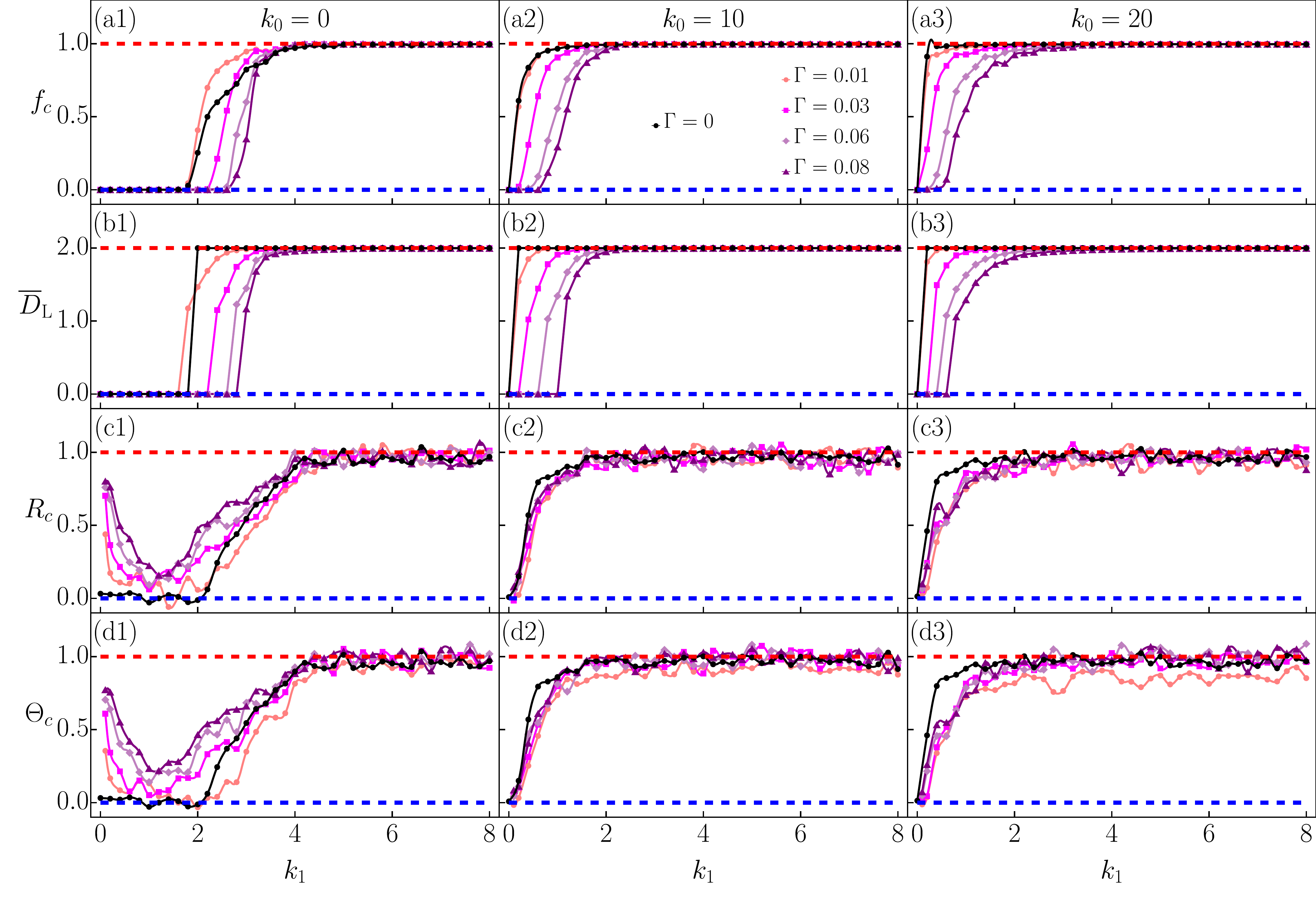}
\caption{Classical and quantum signatures of chaos for the kicked top with low dissipation ($\Gamma<0.1$). [(a1)-(a3)] Fraction of chaotic initial conditions in Eq.~(15) of the main text and [(b1)-(b3)] Average Lyapunov dimension in Eq.~(16) of the main text as a function of the kick strength $k_{1}$. Normalized averages [(c1)-(c3)] $R_{c}$ and [(d1)-(d3)] $\Theta_{c}$ in Eq.~\eqref{eq:NormalizedSpectralRatio} as a function of the kick strength. Each column represents a different case: [(a1)-(d1)] $k_{0}=0$, [(a2)-(d2)] $k_{0}=10$, and [(a3)-(d3)] $k_{0}=20$. In panels (a1)-(a3) and (b1)-(b3), we evolve a set of 1250 initial conditions until $10^{3}$ periods for each kick strength and low dissipation strengths $\Gamma=0.01,0.03,0.06,0.08$. The black line represents the chaos fraction in Eq.~\eqref{eq:ChaosFraction} for the isolated case ($\Gamma=0$). In panels (a1)-(a3), the horizontal blue (red) dashed line represents the limit of a classical system with simple (chaotic) attractors. In panels (b1)-(b3), the horizontal blue (red) dashed line represents the Lyapunov dimension of a point (surface). In panels (c1)-(c3) and (d1)-(d3), we chose a system size $j=80$ and use the positive parity sector of the Floquet operator for each kick strength and low dissipation strengths $\Gamma=0.01,0.03,0.06,0.08$. The black line represents the normalized spectral ratio in Eq.~\eqref{eq:SpectralRatio} for the isolated case ($\Gamma=0$). The horizontal blue (red) dashed line represents the limit of a quantum system described by 2D Poisson (GinUE) statistics. We use $p=2$ in all panels.}
\label{fig:WeakDissipation}
\end{figure}

Additionally, Figs.~\ref{fig:BifurcationDiagram}(c1)-\ref{fig:BifurcationDiagram}(c3) and Figs.~\ref{fig:BifurcationDiagram}(d1)-\ref{fig:BifurcationDiagram}(d3) show the Lyapunov dimension and the Hausdorff dimension, respectively, as a function of dissipation strength. In these figures, we observe that both dimensions behave similarly, becoming zero in dissipation regions where the maximum Lyapunov exponent is negative and point attractors emerge. In contrast, for dissipation strengths within the range $0<\Gamma\leq0.4$, the maximum Lyapunov exponent stays positive, and both the Lyapunov and Hausdorff dimensions approximate the isolated value $\overline{D}_{\text{L}}=\overline{D}_{\text{H}}=2$. The value $\Gamma=0.4$ is indicated by an orange dotted line in all figures.

\subsection{Correspondence principle for low dissipation}

We investigate the correspondence principle in the kicked top system with low dissipation. In our classical analysis, we calculate the fraction of chaotic initial conditions, denoted as $f_{c}$, as described in Eq.~(15) of the main text. We also compute the Lyapunov dimension $D_{\text{L}}$, as outlined in Eq.~(16) of the main text. For the quantum analysis, we use the averages $\langle r \rangle$ and $-\langle \cos\theta \rangle$ from the complex spectral ratio presented in Eq.~(14) of the main text to define normalized quantities
\begin{equation}
    \label{eq:NormalizedSpectralRatio}
    R_{c} = \frac{\langle r\rangle - \langle r\rangle_{\text{2DP}}}{\langle r\rangle_{\text{GinUE}} - \langle r\rangle_{\text{2DP}}} , \hspace{0.5in} \Theta_{c} = \frac{\langle \cos\theta\rangle - \langle \cos\theta\rangle_{\text{2DP}}}{\langle \cos\theta\rangle_{\text{GinUE}} - \langle \cos\theta\rangle_{\text{2DP}}} ,
\end{equation}
where $R_{c}=\Theta_{c}=1$ identifies a quantum system following GinUE statistics, while $R_{c}=\Theta_{c}=0$ represents a quantum system described by 2D Poisson statistics. 

In Fig.~\ref{fig:WeakDissipation}, we illustrate the relationship between the classical and quantum signatures of chaos in the kicked top system with low dissipation, as a function of kick strength. We focus on small dissipation values ($\Gamma < 0.1$) and consider the previous cases of $k_{0}=0,10,20$. Figure~\ref{fig:WeakDissipation}(a1) depicts the fraction of chaotic initial conditions for $k_{0} = 0$, while Fig.~\ref{fig:WeakDissipation}(a2) corresponds to $k_{0} = 10$, and Fig.~\ref{fig:WeakDissipation}(a3) relates to $k_{0} = 20$. We show the correspondence between $f_{c}$ for different dissipation strengths and the chaos fraction $\mu_{c}$ in Eq.~\eqref{eq:ChaosFraction} for the isolated system ($\Gamma = 0$, black line). In addition, Figures~\ref{fig:WeakDissipation}(b1)-\ref{fig:WeakDissipation}(b3) present the corresponding average Lyapunov dimension for each previous case, including a comparison with the isolated case. All figures exhibit a similar pattern across different dissipation strengths and tend to align with the isolated case, showing deviations in the transition from regular to chaotic motion as dissipation increases.

Furthermore, Figs.~\ref{fig:WeakDissipation}(c1)-\ref{fig:WeakDissipation}(c3) illustrate the normalized average $R_{c}$ as defined in Eq.~\eqref{eq:NormalizedSpectralRatio} for the cases where $k_{0}=0,10,20$. In addition, Figs.~\ref{fig:WeakDissipation}(d1)-\ref{fig:WeakDissipation}(d3) present the normalized average $\Theta_{c}$ also outlined in Eq.~\eqref{eq:NormalizedSpectralRatio}. We establish a connection between these normalized averages for different dissipation strengths and the normalized spectral ratio $r_{c}$ given in Eq.~\eqref{eq:SpectralRatio} for the isolated system ($\Gamma = 0$, black line). For the case where $k_{0}=0$, we observe a discrepancy among different dissipation strengths in the weak kick region, with the trends only converging to the isolated scenario in the strong kick region. In contrast, the other cases of $k_{0}=10$ and $k_{0}=20$ exhibit a more consistent trend, closely aligning with the isolated case across all kick strengths. We observe an overall agreement between the classical and quantum tests for very low dissipation strengths and a strong value of the parameter $k_{0}$, which facilitates an overall quantum transition regardless of the dissipation strength.

\section{Dissipation process in the kicked top}
\label{sec:DissipationProcess}

\subsection{Dissipation decoupled from unitary evolution}

We review a method for analyzing a dissipative process that is decoupled from unitary evolution, originally presented in Ref.~\cite{Grobe1987}, which addresses the impact of dissipation on quantum evolution of the kicked top. In this framework, dissipation is introduced into the isolated system after its unitary evolution has taken place. This approach has proven advantageous for theoretical derivations and for inspiring experimental setups~\cite{Braun1998a,Braun1998b,Braun1999,Braun1999PhysD}. Here, we show that this method yields classical results equivalent to those obtained with the exact approach described in the main text, and it can be utilized for further investigations.

The procedure to obtain a dissipation process independent of the unitary evolution consists on splitting the dissipative Floquet operator $\hat{D}$ as follows~\cite{Grobe1987}
\begin{equation}
    \label{eq:DissipationKick}
    \hat{\rho}_{n+1} = \hat{D}\hat{\rho}_{n} = e^{- i \hat{L}_{1}}e^{\left(\hat{\Lambda} - i \hat{L}_{0}\right)} \hat{\rho}_{n} \approx e^{- i \hat{L}_{1}}\left(e^{\hat{\Lambda}}e^{- i \hat{L}_{0}}\right) \hat{\rho}_{n} = e^{\hat{\Lambda}}\left(e^{- i \hat{L}_{0}}e^{- i \hat{L}_{1}}\right) \hat{\rho}_{n} = e^{\hat{\Lambda}}\left(\hat{F}\hat{\rho}_{n}\hat{F}^{\dagger}\right) ,
\end{equation}
where the dissipation operator $\hat{\Lambda}$ is given in Eq.~(2) of the main text, $\hat{L}_{0,1}\hat{\rho}=[\hat{H}_{0,1},\hat{\rho}]$, and $\hat{F}=e^{-i\hat{H}_{1}}e^{-i\hat{H}_{0}}$ is the Floquet operator of the kicked top in absence of dissipation. In Eq.~\eqref{eq:DissipationKick}, we identify three main processes involved. The first two processes are reversible and describe unitary evolution; these are the precessional and torsional motions, represented by the superoperators $\hat{L}_{0}$ and $\hat{L}_{1}$, respectively. The third process is irreversible and occurs after the unitary evolution, represented by the dissipation operator $\hat{\Lambda}$.

The superoperators $\hat{\Lambda}$ and $\hat{L}_{0}$ do not commute, $[\hat{\Lambda},\hat{L}_{0}]\neq0$. Therefore, Eq.~\eqref{eq:DissipationKick} is only applicable for low dissipation strengths and provides an approximate description of the system. However, as noted in Refs.~\cite{Peplowski1991,Iwaniszewski1995}, incorporating a pumping process into the dissipation operator allows Eq.~\eqref{eq:DissipationKick} to be valid across all dissipation strengths. The explicit form of such an operator is as follows
\begin{equation}
    \hat{\Lambda}_{p}\hat{\rho} = \frac{\Gamma}{2j}\left(2\hat{J}_{-}\hat{\rho}\hat{J}_{+} - \{\hat{J}_{+}\hat{J}_{-},\hat{\rho}\}\right) + \frac{\Gamma_{p}}{2j}\left(2\hat{J}_{+}\hat{\rho}\hat{J}_{-} - \{\hat{J}_{-}\hat{J}_{+},\hat{\rho}\}\right) ,
\end{equation}
where $\Gamma$ ($\Gamma_{p}$) is the dissipation (pumping) strength and $[\hat{\Lambda}_{p},\hat{L}_{0}]=0$.

\subsection{Map of dissipative classical motion}

When considering dissipation independent of unitary dynamics, we can derive a map that illustrates the classical motion of the system. This is accomplished by solving the equations of motion for the kicked top specifically for the dissipation process. In this context, we set $p=k_{0}=k_{1}=0$ while keeping $\Gamma\neq0$ in Eqs.~(5)-(7) of the main text, which leads to the following system of equations
\begin{align}
    \dot{J}_{x} = & \Gamma J_{x}J_{z} , \\
    \dot{J}_{y} = & \Gamma J_{y}J_{z} , \\
    \dot{J}_{z} = & - \Gamma \left(J_{x}^{2} + J_{y}^{2}\right) .
\end{align}
Using the constriction for the variables on the Bloch sphere, $J_{x}^{2} + J_{y}^{2} + J_{z}^{2} = 1$, we arrive at the solutions
\begin{equation}
    \left(\begin{array}{c}
       J_{x} \\
       J_{y} \\
       J_{z}
    \end{array}\right) = \frac{1}{\cosh(\Gamma) - \sinh(\Gamma)J_{z}^{0}} \left(\begin{array}{c}
       J_{x}^{0} \\
       J_{y}^{0} \\
       \cosh(\Gamma)J_{z}^{0} - \sinh(\Gamma)
    \end{array}\right)
\end{equation}
with initial conditions $J_{x,y,z}^{0}$. The last expressions are independent of the unitary evolution and allow us to define a kick-to-kick stroboscopic description including dissipation as
\begin{align}
    \label{eq:JxClassicalKick}
    \widetilde{J}_{x}^{n+1} = & \frac{J_{x}^{n+1}}{\cosh(\Gamma) - \sinh(\Gamma)J_{z}^{n+1}} , \\
    \label{eq:JyClassicalKick}
    \widetilde{J}_{y}^{n+1} = & \frac{J_{y}^{n+1}}{\cosh(\Gamma) - \sinh(\Gamma)J_{z}^{n+1}} , \\
    \label{eq:JzClassicalKick}
    \widetilde{J}_{z}^{n+1} = & \frac{\cosh(\Gamma)J_{z}^{n+1} - \sinh(\Gamma)}{\cosh(\Gamma) - \sinh(\Gamma)J_{z}^{n+1}} ,
\end{align}
where $J_{x,y,z}^{n+1} = R(\Omega_{1}^{n})R(\Omega_{0}^{n})R(p)J_{x,y,z}^{n} $ is the map of isolated classical motion given in Eq.~\eqref{eq:IsolatedClassicalMap}. If we set $\Gamma=0$ in Eqs.~\eqref{eq:JxClassicalKick}-\eqref{eq:JzClassicalKick}, we recover the isolated dynamics, $\widetilde{J}_{x,y,z}^{n+1} = J_{x,y,z}^{n+1}$, as expected.

\subsection{Correspondence between coupled and decoupled dissipation}

\begin{figure}[ht]
\centering
\includegraphics[width=0.95\textwidth]{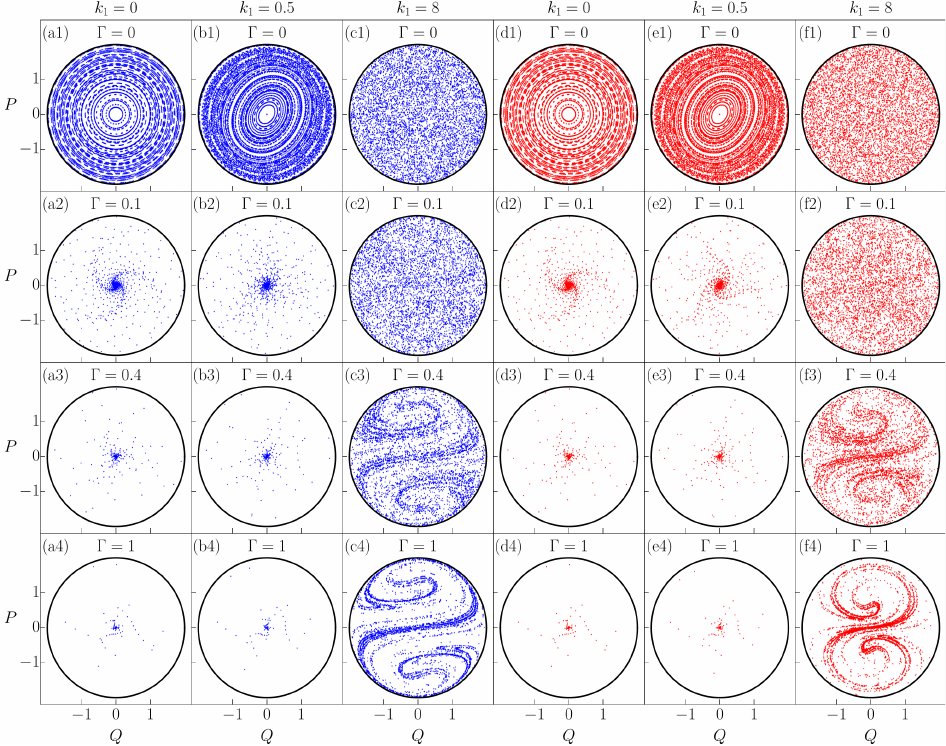}
\caption{Comparison of the classical evolution of the kicked top for coupled and decoupled dissipation. Blue columns [(a1)-(a4)], [(b1)-(b4)], and [(c1)-(c4)] represent the classical map presented in Eqs.~(11)-(16) in the main text for coupled dissipation, while red columns [(d1)-(d4)], [(e1)-(e4)], and [(f1)-(f4)] represent the classical map given in Eqs.~\eqref{eq:JxClassicalKick}-\eqref{eq:JzClassicalKick} for decoupled dissipation. Each column identifies a different kick strength: [(a1)-(a4)] $k_{1}=0$, [(b1)-(b4)] $k_{1}=0.5$, [(c1)-(c4)] $k_{1}=8$, [(d1)-(d4)] $k_{1}=0$, [(e1)-(e4)] $k_{1}=0.5$, and [(f1)-(f4)] $k_{1}=8$. In contrast, each row identifies a different dissipation strength: [(a1)-(f1)] $\Gamma=0$, [(a2)-(f2)] $\Gamma=0.1$, [(a3)-(f3)] $\Gamma=0.4$, and [(a4)-(f4)] $\Gamma=1$. In each panel, we evolve a set of 300 initial conditions until $10^{3}$ periods for each kick and dissipation strengths. The black circle represents the boundary of the initial phase space. System parameters: $p=2$ and $k_{0}=0$.}
\label{fig:PoincareSection_ko0}
\end{figure}

Here, we show the classical correspondence between the dissipation description outlined in Eqs.~(5)-(7) of the main text and Eqs.~\eqref{eq:JxClassicalKick}-\eqref{eq:JzClassicalKick}. We compute Poincar\'e sections in the position and momentum variables $(Q,P)$ for various kick and dissipation strengths. Specifically, we select three different values for the kick strength, which exhibit integrable ($k_{1}=0$), mixed ($k_{1}=0.5$), and chaotic ($k_{1}=8$) dynamics in the absence of dissipation. Next, we analyze the classical evolution under conditions of weak and high dissipation to identify the presence of simple and chaotic attractors. The results are presented in Fig.~\ref{fig:PoincareSection_ko0} for $k_{0}=0$, Fig.~\ref{fig:PoincareSection_ko10} for $k_{0}=10$, and Fig.~\ref{fig:PoincareSection_ko20} for $k_{0}=20$. 

As expected, both descriptions yield the same results for the isolated case ($\Gamma=0$). When dissipation is considered, the descriptions remain approximately equivalent for dissipation strengths in the range of $0<\Gamma\leq0.4$. While higher dissipation strengths are not permissible in the quantum case, they are well-defined in classical terms. However, as we increase the dissipation strength ($\Gamma > 0.4$) for strong kick strengths, noticeable differences emerge in the structure of chaotic attractors. This can be observed in the following pairs of panels: Fig.~\ref{fig:PoincareSection_ko0}(c4) and Fig.~\ref{fig:PoincareSection_ko0}(f4) for $k_{0}=0$; Fig.~\ref{fig:PoincareSection_ko10}(c4) and Fig.~\ref{fig:PoincareSection_ko10}(f4) for $k_{0}=10$; and Fig.~\ref{fig:PoincareSection_ko20}(c4) and Fig.~\ref{fig:PoincareSection_ko20}(f4) for $k_{0}=20$.

\begin{figure}[ht]
\centering
\includegraphics[width=0.95\textwidth]{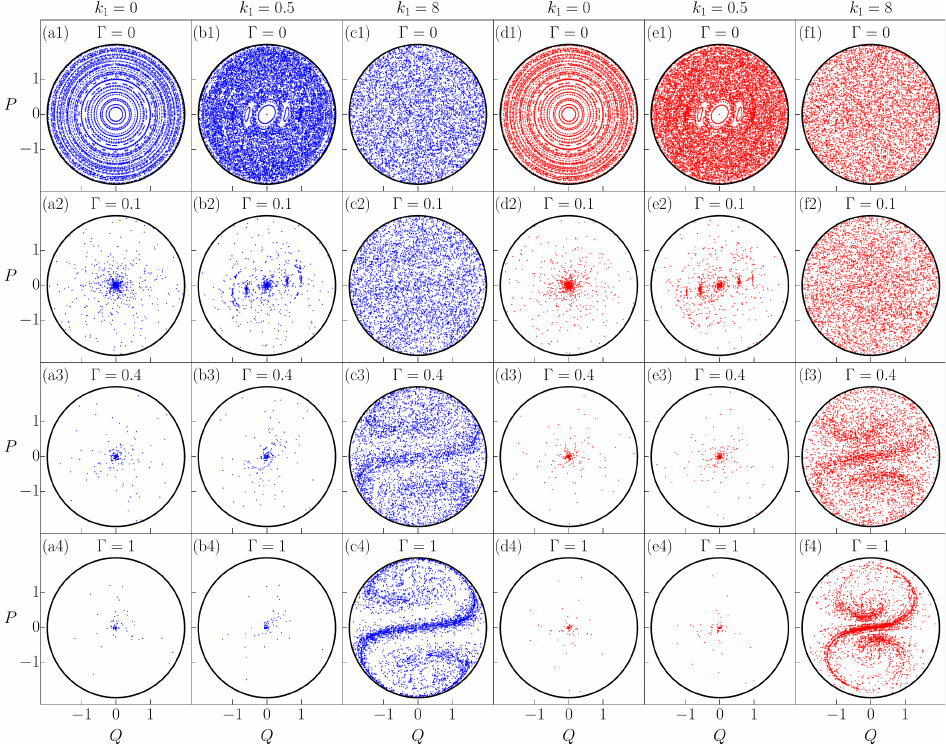}
\caption{Comparison of the classical evolution of the kicked top for coupled and decoupled dissipation. Blue columns [(a1)-(a4)], [(b1)-(b4)], and [(c1)-(c4)] represent the classical map presented in Eqs.~(11)-(16) in the main text for coupled dissipation, while red columns [(d1)-(d4)], [(e1)-(e4)], and [(f1)-(f4)] represent the classical map given in Eqs.~\eqref{eq:JxClassicalKick}-\eqref{eq:JzClassicalKick} for decoupled dissipation. Each column identifies a different kick strength: [(a1)-(a4)] $k_{1}=0$, [(b1)-(b4)] $k_{1}=0.5$, [(c1)-(c4)] $k_{1}=8$, [(d1)-(d4)] $k_{1}=0$, [(e1)-(e4)] $k_{1}=0.5$, and [(f1)-(f4)] $k_{1}=8$. In contrast, each row identifies a different dissipation strength: [(a1)-(f1)] $\Gamma=0$, [(a2)-(f2)] $\Gamma=0.1$, [(a3)-(f3)] $\Gamma=0.4$, and [(a4)-(f4)] $\Gamma=1$. In each panel, we evolve a set of 300 initial conditions until $10^{3}$ periods for each kick and dissipation strengths. The black circle represents the boundary of the initial phase space. System parameters: $p=2$ and $k_{0}=10$.}
\label{fig:PoincareSection_ko10}
\end{figure}

\begin{figure}[ht]
\centering
\includegraphics[width=0.95\textwidth]{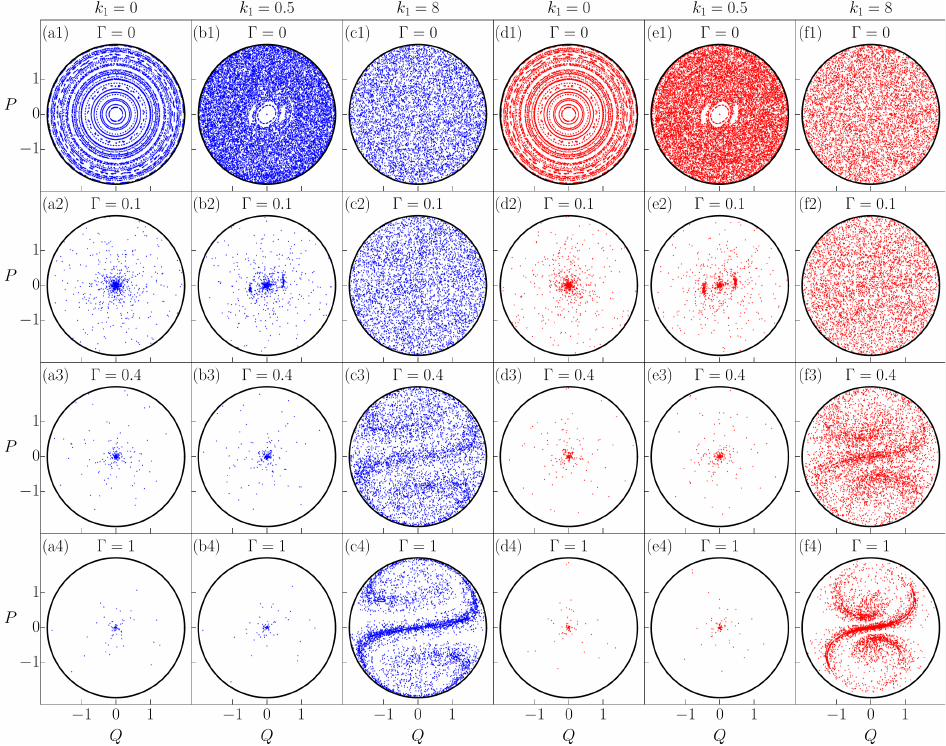}
\caption{Comparison of the classical evolution of the kicked top for coupled and decoupled dissipation. Blue columns [(a1)-(a4)], [(b1)-(b4)], and [(c1)-(c4)] represent the classical map presented in Eqs.~(11)-(16) in the main text for coupled dissipation, while red columns [(d1)-(d4)], [(e1)-(e4)], and [(f1)-(f4)] represent the classical map given in Eqs.~\eqref{eq:JxClassicalKick}-\eqref{eq:JzClassicalKick} for decoupled dissipation. Each column identifies a different kick strength: [(a1)-(a4)] $k_{1}=0$, [(b1)-(b4)] $k_{1}=0.5$, [(c1)-(c4)] $k_{1}=8$, [(d1)-(d4)] $k_{1}=0$, [(e1)-(e4)] $k_{1}=0.5$, and [(f1)-(f4)] $k_{1}=8$. In contrast, each row identifies a different dissipation strength: [(a1)-(f1)] $\Gamma=0$, [(a2)-(f2)] $\Gamma=0.1$, [(a3)-(f3)] $\Gamma=0.4$, and [(a4)-(f4)] $\Gamma=1$. In each panel, we evolve a set of 300 initial conditions until $10^{3}$ periods for each kick and dissipation strengths. The black circle represents the boundary of the initial phase space. System parameters: $p=2$ and $k_{0}=20$.}
\label{fig:PoincareSection_ko20}
\end{figure}

\end{widetext}

\end{document}